\documentclass[ALICE,manyauthors]{cernphprep}

\newcommand{\pp}{p\kern-0.05em p}
\newcommand{\PbPb}{\ensuremath{\mbox{Pb--Pb}}}
\newcommand{\GeVc}{\ensuremath{\mathrm{GeV}\kern-0.05em/\kern-0.02em c}}
\newcommand{\MeVc}{\ensuremath{\mathrm{MeV}\kern-0.05em/\kern-0.02em c}}
\newcommand{\sqrts}{\ensuremath{\sqrt{s_{\mathrm{NN}}}}}
\newcommand{\pT}{\ensuremath{p_{\mathrm{T}}}}

\newcommand{\pTlead}{\ensuremath{p_{\mathrm{T}}^{\mathrm{lead,ch}}}}
\newcommand{\pTtrack}{\ensuremath{p_{\mathrm{T,track}}}}
\newcommand{\pTclus}{\ensuremath{p_{\mathrm{T,cluster}}}}

\newcommand{\Eclus}{\ensuremath{E_{\mathrm{cluster}}}}
\newcommand{\Ehadcorr}{\ensuremath{E_{\mathrm{cluster}}^{\mathrm{hadcorr}}}}
\newcommand{\Enonlincorr}{\ensuremath{E_{\mathrm{cluster}}^{\mathrm{nonlincorr}}}}

\newcommand{\etajet}{\ensuremath{\eta_{\mathrm{jet}}}}
\newcommand{\pTjet}{\ensuremath{p_{\mathrm{T,jet}}}}
\newcommand{\pTreco}{\ensuremath{p_{\mathrm{T,jet}}^{\mathrm{reco}}}}
\newcommand{\pTraw}{\ensuremath{p_{\mathrm{T,jet}}^{\mathrm{raw}}}}
\newcommand{\pTtrue}{\ensuremath{p_{\mathrm{T,jet}}^{\mathrm{true}}}}
\newcommand{\pTminreco}{\ensuremath{p_{\mathrm{T,jet}}^{\mathrm{reco,min}}}}
\newcommand{\pTmaxreco}{\ensuremath{p_{\mathrm{T,jet}}^{\mathrm{reco,max}}}}
\newcommand{\pTmin}{\ensuremath{p_{\mathrm{T,jet}}^{\mathrm{min}}}}
\newcommand{\pTmax}{\ensuremath{p_{\mathrm{T,jet}}^{\mathrm{max}}}}
\newcommand{\deltaPt}{\ensuremath{\delta_{\pT}}}
\newcommand{\Ecell}{\ensuremath{E_{\mathrm{cell}}}}
\newcommand{\Eseed}{\ensuremath{E_{\mathrm{seed}}}}
\newcommand{\Raa}{\ensuremath{R_{\mathrm{AA}}}}
\newcommand{\Ncoll}{\ensuremath{N_{\mathrm{coll}}}}

\newcommand{\Taa}{\ensuremath{\left<T_{\mathrm{AA}}\right>}}

\usepackage[comma,square,numbers,sort&compress]{natbib}
\usepackage{hyperref}
\usepackage{lineno}
\usepackage[T1]{fontenc}

\begin{document}%

\begin{titlepage}
\PHyear{2019}
\PHnumber{200}      
\PHdate{18 September}  
%

\title{Measurements of inclusive jet spectra in \pp{} and central \PbPb{} collisions at $\mathbf{\sqrt{\textit{s}_{\mathrm{\mathbf{NN}}}}} \mathbf{\;= 5.02}$ TeV}
\ShortTitle{Measurements of jet spectra in \pp{} and \PbPb{} collisions at $\sqrts=5.02$ TeV}   

\Collaboration{ALICE Collaboration\thanks{See Appendix~\ref{app:collab} for the list of collaboration members}}
\ShortAuthor{ALICE Collaboration} 

\begin{abstract}
This article reports measurements of the \pT{}-differential inclusive jet cross-section in \pp{} collisions at {$\sqrt{s}=5.02$ TeV} and the \pT{}-differential 
inclusive jet yield in \PbPb{} 0--10\% central collisions at $\sqrts=5.02$ TeV. 
Jets were reconstructed at mid-rapidity with the ALICE tracking detectors and electromagnetic calorimeter using the anti-$k_\mathrm{T}$ algorithm.
For \pp{} collisions, we report jet cross-sections for jet resolution parameters $R=0.1-0.6$ over the range $20<\pTjet<140$ \GeVc, as well as the jet cross-section ratios of different $R$,
and comparisons to two next-to-leading-order (NLO)-based theoretical predictions.
For \PbPb{} collisions, we report the $R=0.2$ and $R=0.4$ jet spectra for $40<\pTjet<140$ \GeVc{} and $60<\pTjet<140$ \GeVc, respectively.
The scaled ratio of jet yields observed in \PbPb{} to \pp{} collisions, \Raa, is constructed, and exhibits strong jet quenching and a clear \pT{}-dependence for $R=0.2$.
No significant $R$-dependence of the jet \Raa{} is observed within the uncertainties of the measurement. 
These results are compared to several theoretical predictions.

\end{abstract}
\end{titlepage}
\setcounter{page}{2}

\section{Introduction}

A deconfined state of strongly interacting matter described by Quantum Chromodynamics (QCD) is produced in ultra-relativistic heavy-ion collisions at
the Relativistic Heavy Ion Collider (RHIC) and the Large Hadron Collider (LHC) \cite{Jacak:2012dx, LHC1review, Braun-Munzinger:2015hba, TheBigPicture, PhenixQGP, StarQGP, PhobosQGP, BrahmsQGP}. 
Numerous observables including high-\pT{} hadron suppression, anisotropic flow, and $J/\psi$ suppression and recombination provide evidence that the hot QCD
state produced in these collisions consists of sub-nucleonic degrees of freedom. 

One of the major strategies to investigate this hot QCD state is the study of jet modification in heavy-ion collisions.
Partons often traverse a significant pathlength of the hot QCD medium, and the effect that the medium has on the resulting jets can be deduced by comparing jet properties in heavy-ion collisions to those in \pp{} collisions.
Since the jet production cross-section can be computed in perturbative QCD, and since jets are sensitive to a wide range of momentum exchanges with the medium, 
jet physics is an appealing tool to investigate the medium at a wide range of resolution scales.

Previous measurements demonstrate suppression of the jet transverse momentum (\pT{}) spectrum in heavy-ion collisions relative to \pp{} collisions scaled by the number of incoherent binary nucleon-nucleon collisions, 
indicating that jets transfer energy to the hot QCD medium \cite{fullJet276, chJet276, hjetPbPb, hjetAuAu, jetRaa276Atlas, atlas502, jetRaa276CMS}. Furthermore, jet substructure measurements indicate that in
heavy-ion collisions, the jet core is more collimated and fragments harder \cite{AliceJetShape}, while at wide angles from the jet axis there
is an excess of soft particles \cite{CMSdijetimbalance, atlasFF502}.
Jet modification in heavy-ion collisions is described by several different theoretical approaches typically based on energy loss via 
medium-induced gluon radiation and elastic scattering \cite[and references therein]{JETcollaboration, ReviewXinNian, ReviewYacine, ReviewMajumder};
however, there remains no clear consensus of the precise nature of the interaction of jets with the medium. 
New measurements of the absolute level of jet suppression and its \pT{}-dependence will directly test models, 
and serve as a key constraint for global analyses of high-\pT{} observables.
Additionally, the evolution of jet suppression with the jet resolution parameter, $R$, can constrain competing effects between
the recovery of out-of-cone radiation and the changing selection of the jet population (such as reduction of the quark/gluon fraction) as $R$ increases \cite{Qiu:2019sfj, JewelMediumResponse, LBTconspiracy}.

The inclusive jet cross-section in \pp{} collisions contains important QCD physics itself.
In recent years, the inclusive jet cross-section in \pp{} collisions was computed at NLO with resummation of logarithms of the jet 
resolution parameter \cite{Dasgupta2015, Dasgupta2016, SiJF, NLLSCET} and threshold logarithms \cite{JointResummation2017, JointResummation2018}, and also to NNLO both with and without the leading color approximation \cite{NNLO, Czakon2019}.
Measurements of the inclusive \pp{} jet cross-section have been made at the SPS \cite{ppUA1, ppUA2}, the Tevatron \cite{ppCDF, ppD0}, RHIC \cite{ppSTAR},
and the LHC \cite{ppALICE502, ppALICE276, ppCMS276TeV, ppCMS7TeV, ppCMS8TeV, ppCMS13TeV, ppATLAS276TeV, ppATLAS7TeV, ppATLAS7TeV-2}, 
and the latest comparisons of these measurements with theoretical predictions demonstrate the importance of 
contributions beyond NLO fixed-order calculations, namely resummations or matched parton showers.
However, the precise contributions of the perturbative aspect of the jet, as well as the hadronization and underlying event (UE) effects, remain under investigation. 
Inclusive jet measurements at low-\pT{} as a function of $R$ (including ratios of jet cross sections, which allow partial cancellation of experimental and theoretical uncertainties)
will help clarify these contributions, and provide tests for both the perturbative and non-perturbative contributions to the inclusive jet cross-section.
Moreover, these measurements can be used to constrain parton distribution functions (PDFs) and the strong coupling constant $\alpha_{\mathrm{s}}$ \cite{Britzger2019, Ball2018, Malaescu2012, ppCMS8TeV, ppATLAS276TeV}.

This article reports measurements of inclusive jet \pT{} spectra in \pp{} and central \PbPb{} collisions at $\sqrts=5.02$ TeV with the ALICE detector.
Jets were reconstructed in the pseudo-rapidity range $|\etajet|<0.7-R$ for jet resolution parameters $R=0.1-0.6$ in \pp{} collisions and $R=0.2$ and $R=0.4$ in \PbPb{} collisions.
In \PbPb{} collisions, we required jets to contain at least one charged track with $\pT>5-7$ GeV$/c$ (depending on the jet $R$) in order to identify hard jet candidates (arising from large momentum-transfer scatterings) in the 
large background from combinatorial jets. In \pp{} collisions, we report the cross-section both with and without this bias. 
The relative jet yields observed in \PbPb{} and \pp{} collisions are reported using their scaled ratio, \Raa, and compared to several theoretical predictions. 
\section{Experimental Setup and Datasets}

The ALICE detector \cite{aliceDetector, alicePerformance} is a dedicated heavy-ion experiment located at the Large Hadron Collider \cite{LHCmachine}. 
The analysis relied on the central tracking system and the electromagnetic calorimeter (EMCal), as well as detectors for event triggering and centrality determination.
The tracking system consists of a six-layer silicon inner tracking system (ITS) with radial distance 3.9\,--\,43\,cm from the beamline,
and a gas time projection chamber (TPC) with radial distance 85\,--\,247\,cm from the beamline. The combined tracking system
spans $\left|\eta\right|<0.9$ and full azimuth, and tracks were measured in the range $150 \; \MeVc < \pTtrack < 100 \; \GeVc$. 
The EMCal consists of a Pb-scintillator sampling calorimeter spanning $\left|\eta\right|<0.7$ and $1.4<\varphi<3.3$ in azimuth, 
located a radial distance 4.36\,m from the beamline \cite{EMCalTDR}. 
It contains 12,288 cells organized in an approximately projective geometry relative to the interaction point.
The Moliere radius of the EMCal is $r_{\mathrm{M}} = 3.2$\,cm, and its cells have a transverse size of approximately $6.0$\,cm$\times6.0$\,cm ($\Delta\eta\times\Delta\varphi \approx 0.014\times0.014$).
Each cell has a depth of 24.6\,cm, corresponding to approximately 20 electromagnetic radiation lengths and one hadronic interaction length. 

The reported \PbPb{} (\pp) data were recorded in 2015 (2017) at $\sqrts=5.02$ TeV. 
The events were collected using a minimum bias (MB)
trigger requiring a coincidence hit in both of the V0 scintillators, located at $2.8<\eta<5.1$ (V0-A) and $-3.7<\eta<-1.7$ (V0-C) \cite{ppXsec}.
An accepted event was required to have a primary vertex successfully reconstructed within $-10\;\mathrm{cm} < z < 10\;\mathrm{cm}$ of the 
interaction point, and to satisfy several vertex quality criteria. 
In \PbPb{} collisions, the centrality was determined using the V0 multiplicities \cite{centrality276, centrality502, Loizides:2017ack}.
Additionally, out-of-bunch pileup was rejected using timing cuts as well as correlating track multiplicities between several subdetectors.
We utilized a sample of approximately 4.6M 0--10\% most central \PbPb{} accepted events (6.0\;$\mu \mathrm{b}^{-1}$) and 760M \pp{} accepted events (15.7\,nb$^{-1}$).

Reconstructed tracks were generally required to include at least one hit in the Silicon Pixel Detector (SPD) comprising the first two layers of the ITS, 
and to have at least 70 TPC space-points and at least 80\% of the geometrically findable space-points in the TPC. 
Tracks without any hits in the SPD, but otherwise satisfying the tracking criteria, were re-fit with a constraint to the primary vertex of the event. 
Including this second class of tracks ensured approximately uniform acceptance in $\varphi$, while preserving similar \pT{} resolution to tracks with SPD hits.
Tracks with $\pTtrack{} >150$\,\MeVc{} were accepted over ${-0.9<\eta<0.9,\;0<\varphi<2\pi}$.
The performance of the detector was estimated with a model of the ALICE detector and its response to particles using GEANT3. 
The tracking efficiency in \pp{} collisions, as estimated by PYTHIA8 Monash 2013 \cite{pythia} and the ALICE GEANT3 detector simulation, is approximately 67\% 
at $\pTtrack=150$\,\MeVc, and rises to approximately 84\% at $\pTtrack=1$ \GeVc{} and remains above 75\% at higher \pT.
Studies of the centrality dependence of the tracking efficiency in a HIJING \cite{hijing} simulation demonstrated that the tracking efficiency 
is approximately 2\% lower in 0-10\% central \PbPb{} collisions compared to \pp{} collisions, independent of \pTtrack.
The momentum resolution $\delta\pT/\pT$ was estimated from the covariance matrix of the track fit \cite{alicePerformance} using PYTHIA8 Monash 2013,
and was approximately 1\% at $\pTtrack=1$ \GeVc{} and 4\% at ${\pTtrack=50 \;\GeVc}$.

Reconstructed EMCal clusters were built by clustering EMCal cells with $\Ecell>100$ MeV around a seed
cell with $\Eseed>300$ MeV, using a clustering algorithm that allows each cluster to have only a single local maximum.
The highest-energy cell in a cluster was required to satisfy a timing cut.
Clusters with large apparent energy but anomalously small number of contributing cells were removed from the analysis, 
since they are believed to be due to interactions of slow neutrons or highly ionizing particles in the avalanche photodiodes \cite{fullJet276}.
The linearity of the energy response of the EMCal was determined from electron test beam data, and a correction of 
about 7\% at $\Eclus = 0.5$ GeV but negligible above ${ \Eclus = 3 \;\mathrm{GeV} }$ was applied to the cluster energies. 
A study using the photon conversion method demonstrated that with this non-linearity correction, the $\pi^0$ mass in Monte Carlo (MC) simulations matches that in \pp{} data within 1\%.
For \pp{} collisions, an additional correction obtained from a photon conversion analysis was used to reduce the small remaining offset of the energy scale in data and MC simulations \cite{pcm}. 
The energy resolution obtained from electron test beam data was about 15\% at $\Eclus = 0.5$ GeV and better than 5\% above ${\Eclus = 3 \;\mathrm{GeV} }$.

Since the jet energy is reconstructed by combining tracks and clusters, one needs to account for the fact that charged particles deposit energy 
in both the tracking system and the EMCal, as in Ref. \cite{ppALICE276}. In particular, all accepted tracks were propagated to the average shower depth of the EMCal, $r = 440$ cm,
and allowed to match geometrically to at most one cluster; clusters were allowed to have multiple matching tracks. 
If a track was matched within \pT-dependent thresholds ranging from ${(\Delta \eta,\Delta \varphi)\approx(0.037, 0.084)}$ at $\pT=0.15 \;\GeVc$ to $(\Delta \eta,\Delta \varphi)\approx(0.010, 0.015)$ at $\pT=100 \;\GeVc$,  
then a hadronic correction was applied to the cluster: $\Ehadcorr = \Enonlincorr - \Delta E$, 
where $\Enonlincorr$ is the non-linearity corrected cluster energy, and $\Delta E = c \sum_{i}p_{\mathrm{i}}^{\mathrm{track}}$, where $ i $ spans all tracks matched to the cluster, 
$p_{\mathrm{i}}^{\mathrm{track}}$ is the track 3-momentum, and $c$ is the speed of light.
After the above cuts and corrections were performed, clusters with $\Ehadcorr > 300\; \mathrm{MeV}$ were accepted.

\section{Jet Reconstruction}

Jets were reconstructed with $R=0.1-0.6$ in \pp{} collisions and $R=0.2,0.4$ in \PbPb{} collisions using the anti-$k_{\mathrm{T}}$ sequential recombination algorithm implemented in FastJet 3.2.1 \cite{antikt, catchment} 
from the combination of charged particle tracks and hadronically corrected EMCal clusters. We used the \pT{} recombination scheme,
assuming EMCal clusters are massless: $\pTraw = \sum_{i} p_{\mathrm{T,track}}^{i} + \sum_{j} p_{\mathrm{T,cluster}}^{j}$, 
where $p_{\mathrm{T,cluster}}=\Ehadcorr / c$. 

In \PbPb{} collisions, we subtracted the average combinatorial background following the approach in Ref. \cite{fullJet276}. 
The background density $\rho$ was determined in each event, and used to subtract the average background from each jet in that event: $\pTreco = \pTraw - \rho A,$ where $A$ is the jet area.
The average background density in 0--10\% central events is typically $\left<\rho\right>\approx 220-280$ \GeVc, corresponding to $\approx 110-140$ \GeVc{} for a $R=0.4$ jet.
In \pp{} collisions, we did not subtract the background due to the underlying event, in order to minimize the model dependence of the measurement.

Jets selected for the measurement were required to satisfy several criteria in order to be accepted:
(i) the center of the jet must be within the fiducial volume of the EMCal, i.e. a distance $\Delta R \equiv \sqrt{\left(\Delta \eta \right)^2 + \left(\Delta \varphi \right)^2}$ from any edge of the EMCal,
(ii) the jet must not contain any tracks with $\pTtrack > 100$ \GeVc,
(iii) in \PbPb{} and applicable \pp{} results, the jet must contain a track with $\pTtrack > 5-7$ GeV/$c$, depending on $R$, and
(iv) in \PbPb{} collisions, the area of the jet must be $A > 0.6 \pi R^{2}$.
The $\pTtrack<100$ GeV/c requirement  removed only a small number of jets at large \pTreco{}, and has negligible bias for the \pTmaxreco{} selected in this analysis.
The leading track requirement introduces a small fragmentation bias in the jet sample, which may lead to a bias in the measured jet suppression. 
This effect is discussed in Section 6, and is estimated to have only a small effect on the reported \Raa{}. 
A larger leading track requirement is needed for larger $R$ since the magnitude of background fluctuations increases with $R$.
The area cut in \PbPb{} collisions was negligible except at very low \pTreco{}, where it rejects combinatorial jets.

In \PbPb{} collisions, local fluctuations in the background smear the reconstructed jet momentum.
To study jet-by-jet fluctuations in the background, we generated a random $\left(\eta,\varphi\right)$ within the fiducial calorimeter acceptance in each event, 
and compared the sum of constituents in a cone of radius $R$ to the expected average background in that cone: $\deltaPt = \sum_{\mathrm{cone}} \left( \pTtrack+ \pTclus \right)-\rho\pi R^{2}$. 
The width of the \deltaPt{} distribution is a measure of the size of the background fluctuations \cite{alicebkg}. 
For $R=0.2$, the standard deviation of the \deltaPt{} distribution is $\sigma_{\deltaPt}=6.5$ \GeVc, which grows to $\sigma_{\deltaPt}=16.1$ \GeVc{} for $R=0.4$.
In the present analysis the \deltaPt{} distributions were not explicitly used except to determine the \pTreco{} range to utilize in the analysis, which is discussed in Section 4.

We evaluated the performance of our jet reconstruction strategy by estimating the mean jet energy scale shift, $\Delta_{\mathrm{JES}}=\left< \left( \pTreco - \pTtrue \right) / \pTtrue \right>$, 
the jet energy resolution, $\mathrm{JER}=\sigma\left(\pTreco \right) \left/ \pTtrue \right.$, and the jet reconstruction efficiency, $\varepsilon_{\mathrm{reco}}$, 
from PYTHIA8 Monash 2013 and the ALICE detector simulation.
Table \ref{table:jetreco} shows approximate values of $\Delta_{\mathrm{JES}},\; \mathrm{JER},\; \varepsilon_{\mathrm{reco}}$ for $R=0.2$ and $R=0.4$ in \pp{} and \PbPb{} collisions. 
The jet energy scale shift is a long-tailed asymmetric distribution due to reconstruction inefficiency (such as tracking inefficiency) \cite{chJet276}, 
and $\Delta_{\mathrm{JES}}$ should be understood only as a rough characterization of this distribution.
When a leading track requirement is imposed, the jet reconstruction efficiency and jet energy scale shift are primarily due to this requirement
in combination with the tracking efficiency.
Note that the \pp{} response approximately, but not exactly, describes the detector effects in jet reconstruction relevant for \PbPb{} collisions.
In \PbPb{} collisions, the jet reconstruction performance (including the effect of background fluctuations) was determined by embedding \pp{} MC events into \PbPb{} data, as described in detail in Section 4.
The JER is approximately constant at $\approx23\%$ above $\pTtrue=60$ \GeVc{} for $R=0.2$, and deteriorates at lower \pTtrue{} due to background fluctuations.
As $R$ increases, the JER deteriorates due to the increased influence of background fluctuations. 

\begin{table}[]
\centering
\caption{Approximate values characterizing the jet reconstruction performance for $R=0.2$ and $R=0.4$ in \pp{} and \PbPb{} collisions.
For cases with a leading track requirement, $\pTlead = 5\;\GeVc$ is used for $R=0.2$ and $\pTlead = 7\;\GeVc$ for $R=0.4$}
\begin{tabular}{ l ccccccc }
\tabularnewline \hline \hline & \multicolumn{2}{c}{\pp{} $(\pTlead > 0\;\GeVc)$} & \multicolumn{2}{c}{\pp{} $(\pTlead > 5/7\;\GeVc)$}  & \multicolumn{2}{c}{$\PbPb{} \;(\pTlead > 5/7\;\GeVc)$}
\tabularnewline \hline \pTjet{} & $20 \;\GeVc$ & $100\;\GeVc$ & $20\;\GeVc$ & $100\;\GeVc$ & $20\;\GeVc$ & $100\;\GeVc$

\tabularnewline \hline $R=0.2$
\tabularnewline  \quad $\Delta_{\mathrm{JES}}$ & --29\% & --30\% & --18\% & --28\% & --23\% & --35\%
\tabularnewline \quad $\mathrm{JER}$ & 27\% & 21\% & 19\% & 19\% & 35\% & 23\%
\tabularnewline \quad $\varepsilon_{\mathrm{reco}}$ & 98\% & 100\% & 86\% & 96\% & 86\% & 96\%

\tabularnewline \hline $R=0.4$
\tabularnewline \quad $\Delta_{\mathrm{JES}}$ & --30\% & --31\% & --14\% & --27\% & --6\% & --33\%
\tabularnewline \quad $\mathrm{JER}$ & 23\% & 18\% & 15\% & 16\% & 77\% & 25\%
\tabularnewline \quad $\varepsilon_{\mathrm{reco}}$ & 99\% & 100\% & 82\% & 92\% & 82\% & 92\%

\tabularnewline \hline 
\end{tabular}
\label{table:jetreco}
\end{table}
 
\section{Corrections}


The reconstructed \pTreco{} spectrum includes fluctuations in the underlying background (in \PbPb{} collisions) and
a variety of detector effects, including tracking inefficiency, missing long-lived neutral particles (n, $\mathrm{K}_{\mathrm{L}}^{0}$), and particle-material interactions.
We therefore deconvoluted the reconstructed jet spectrum with a response matrix (RM) describing the correlation between \pTreco{} and \pTtrue{}
in order to recover the ``truth"-level jet spectrum at the hadron-level.

In \pp{} collisions, we generated a RM using PYTHIA8 Monash 2013 with the full GEANT3 ALICE detector simulation, based on the detector performance in the relevant 2017 \pp{} data-taking period.
In \PbPb{} collisions, we generated a RM by embedding PYTHIA events (with detector simulation based on the detector performance in the 2015 \PbPb{} data-taking period) into \PbPb{} data after the detector-level reconstruction was run individually on both. 
The set of tracks in the ``hybrid" event was taken as the sum of all tracks in both events individually, 
while the set of EMCal clusters were re-clustered from a combined pool of cells from both events.
This embedding-based approach, which uses real background, ensures that the detector response accurately reflects the \PbPb{} response of the calorimeter, 
including particle overlaps in the calorimeter as well as the \PbPb{} particle composition, 
and ensures the effect of the hadronic correction is equivalent in data and in the response.
Moreover, it ensures that the correlation between the local background and the reconstructed jet due to local detector inefficiencies is accounted for.

The truth-level jet was constructed from the primary particles of the PYTHIA event, defined as all particles with a proper decay length longer 
than 1\,cm, excluding daughters of these particles \cite{primaryParticleALICE}. 
We correct the jet \pT{} to include the ``missing" long-lived neutral particles.

The detector-level jet in \pp{} collisions was constructed from the PYTHIA tracks and clusters at detector level.
In \PbPb{} collisions, the detector-level jet was constructed from the ``hybrid" event consisting of both PYTHIA and \PbPb{} tracks and clusters at detector level.
To account for the decreased tracking efficiency in \PbPb{} collisions, we randomly rejected 2\% of the PYTHIA tracks in the \PbPb{} case, independent of \pT{}.
The average combinatorial background was subtracted as in 0-10\% central \PbPb{} data: we computed the event-by-event 
$\rho_{\mathrm{charged}}$ using only \PbPb{} tracks, and we applied the background scale factor obtained in \PbPb{} data;
we assume that the combinatorial background from the \pp{} event is negligible.

In order to fill the RM, we matched truth-level jets to detector-level jets by a geometrical matching procedure.
In \pp{} collisions, if an accepted detector-level jet and an accepted PYTHIA jet were within $\Delta R<0.6R$,  
and they were both the closest jets to each other, then the jets were matched, and contribute to the RM.
In \PbPb{} collisions, if an accepted hybrid jet and an accepted PYTHIA jet were within $\Delta R<1.5R$, 
and they were both the closest jets to each other, then the jets were matched, and contribute to the RM.
The leading track requirement nullifies the need in \PbPb{} collisions for further criteria such as a shared momentum fraction requirement in order to generate accurate matches.
The RM was generated with 5 \GeVc{} bin widths for \pTreco{} and 10 \GeVc{} widths for \pTtrue{},
and was normalized so as to preserve the number of jets upon unfolding.

To perform the deconvolution, we employed the SVD unfolding algorithm \cite{svd} using the RooUnfold package \cite{roounfold}.
The regularization parameter $k$ suppresses high-frequency variations in the unfolded result, and was selected by examining the so-called $d$-vector distribution.
Statistical uncertainties were computed according to MC pseudo-experiments within RooUnfold.
The reconstructed spectrum was input to the unfolding procedure over a fixed window of 
$\pTreco \in\left[\pTminreco, \pTmaxreco \right]$, as illustrated in Table \ref{table:ptranges}.
In \PbPb{} collisions, each of these \pTminreco{} corresponds to $\approx 2-3 \times \sigma_{\deltaPt}$, which, in combination with the leading charged hadron requirement, 
results in a sample largely free of combinatorial jets.  A larger value of \pTminreco{} was used in \PbPb{} collisions in order to minimize the impact of the combinatorial background, 
which can de-stabilize the unfolding process. Any residual combinatorial jets will still be unfolded to low \pT{} by the RM.
Since truncating the RM in \pTreco{} loses the information of the fraction of truth-level jets that migrate outside of the measured detector-level window,
we corrected for this kinematic efficiency. The unfolded result is then reported in a range over which the input data provides meaningful constraints; 
that is, a region unaffected by combinatorial jets, and where the kinematic efficiency is larger than approximately 80\%. 

\begin{table}[]
\centering
\caption{Minimum and maximum reconstructed jet \pT{} used in the analysis as input to the deconvolution procedure.}
\begin{tabular}{ l ccccccc }
\tabularnewline \hline \hline & \multicolumn{2}{c}{\pp{} \;(\GeVc)} & \multicolumn{2}{c}{\PbPb{} \;(\GeVc)}
\tabularnewline \hline & \pTminreco & \pTmaxreco & \pTminreco & \pTmaxreco
\tabularnewline \hline \quad $R=0.2$ & 7 & 130 & 20 & 120
\tabularnewline \quad $R=0.4$ & 10 & 130 & 35 & 120
\tabularnewline \hline 
\end{tabular}
\label{table:ptranges}
\end{table}

We corrected the unfolded spectrum for the fact that the jet finding procedure failed to reconstruct a certain fraction of jets.
We computed the jet reconstruction efficiency as:
\[
 \varepsilon_{\mathrm{reco}} \left(\pTtrue \right)=N_{\mathrm{matched}}\left(\pTtrue \right) / N_{\mathrm{truth}} \left(\pTtrue \right),
 \]
where $N_{\mathrm{matched}}$ is the number of accepted detector-level jets matched to PYTHIA truth-level jets out of $N_{\mathrm{truth}}$ accepted truth-level jets.
In order that $\varepsilon_{\mathrm{reco}}$ also includes the false positive rate of accepted detector-level jets that have no matching truth-level jet 
(which can occur if the truth-level jet was generated slightly outside of our geometrical acceptance), 
the numerator also contains matches to truth-level jets outside of the EMCal fiducial acceptance.
Note that $\varepsilon_{\mathrm{reco}}$ does not explicitly include the bias of the leading charged hadron requirement, but only the probability to reconstruct an accepted
jet given a truth-level jet satisfying the leading charged hadron requirement (when applicable).
In order for $\varepsilon_{\mathrm{reco}}$ to be the jet reconstruction efficiency, the jet matching efficiency must be 100\%.
However, in the \PbPb{} embedding environment, this is difficult to achieve, since some criteria need to be imposed to 
suppress combinatorial jets (in our case, the leading track requirement). Therefore, in the \PbPb{} case we used the jet reconstruction efficiency as determined 
from a \pp{} simulation alone (with 2\% reduced tracking efficiency).

The unfolded solution was verified to be mathematically robust by performing a re-folding test and a ``self-closure" test.
The re-folding test consisted of generating a RM (from half of the MC data sample runs) and unfolding the measured distribution, 
then applying a RM (from the other half of the MC data sample) to the unfolded result, and comparing the re-folded solution to the measured distribution.
The self-closure test consisted of taking the matched detector-level jet spectrum in the full embedded sample, and smearing each data
point with a Gaussian according to the statistical uncertainties of the measured data. This spectrum was then unfolded using the RM, and
compared the result to the truth-level PYTHIA jet spectrum. In both cases, consistency was achieved within statistical uncertainties.

In \PbPb{} collisions, the unfolded solution is verified to be physically correct by a thermal model closure test similar to that in Ref. \cite{fullJet276}.
The closure test consisted of performing the entire analysis on ``hybrid" events containing a PYTHIA event and a thermal background, 
in which ``hybrid"  jets were clustered from the combination of PYTHIA detector-level particles and thermal background particles. 
The background was modeled by generating $N$ particles from a Gaussian, with \pT{} taken from a Gamma distribution, $f_{\Gamma}\left(\pT ; \beta \right) \sim \pT e^{-\pT / \beta}$, 
where the free parameters $\overline{N},\sigma_{N},\beta$ were fixed to roughly fit the \deltaPt{} distribution in 0--10\% \PbPb{} data.
The test consisted of constructing the hybrid detector-level jet spectrum, building the RM, 
and unfolding the hybrid jets -- and comparing the spectrum to the truth-level PYTHIA spectrum. 
Since the background does not have any jet component, this test is able to verify whether the analysis procedure indeed recovers the jet spectrum, 
and is not contaminated by combinatorial jets. 
These tests validated the analysis procedure within approximately 5\% for $R=0.2$ with $\pTlead=5$ GeV/$c$, and $R=0.4$ with $\pTlead=7$ GeV/$c$.
 
\section{Systematic Uncertainties}


Following Ref. \cite{fullJet276}, we categorized two classes of systematic uncertainties: correlated uncertainties and shape uncertainties.
Correlated uncertainties encompass detector effects such as uncertainty on the tracking efficiency and uncertainty on the 
EMCal response, which are approximately fully positively correlated among all \pTjet{} bins. 
Shape uncertainties refer to systematic unfolding uncertainties, which alter the shape of the final \pTjet{} spectrum. 
The dominant systematic uncertainties in this analysis are the uncertainty in the tracking efficiency and the systematic uncertainty in the unfolding procedure.
Note that in general the following uncertainties describe uncertainties on the jet yield, not on the jet \pT{} scale.

\subsection{Correlated uncertainties}

The dominant correlated uncertainty is the uncertainty on the modeling of the tracking efficiency, since correcting for unmeasured tracks
has a major effect on the unfolding procedure. For the track selection described in Section 2, the uncertainty on the tracking efficiency is
approximately 4\%, as estimated from variation in the track selection parameters and variation in the ITS-TPC matching requirements. 
In order to assign a systematic uncertainty to the final result, we constructed a RM using the same techniques as for the final result except 
with an additional 4\% of PYTHIA tracks randomly rejected in jet finding (for \PbPb{}, this is in addition to the 2\% rejection used for the main result). 
The jet reconstruction efficiency was also computed with this extra 4\% suppression applied.
This modified RM was then used to unfold the same measured spectrum as used for the main result. This varied result was corrected
for the jet reconstruction efficiency, and compared to the main result, with the differences in each bin taken as the uncertainty.
Additionally, the uncertainty due to the tracking \pT{} resolution was approximately 1\%.

\begin{table}[!b]
\centering
\caption{Summary of correlated systematic uncertainties on the \pp{} jet spectra without a leading track bias, for select $R$.
The columns \pTmin{} and \pTmax{} are the uncertainties at the minimum and maximum \pTjet{} bin.}
\begin{tabular}{ l ccccccc }
\tabularnewline \hline \quad & \multicolumn{6}{c}{Relative uncertainty (\%)}

\tabularnewline \hline \hline \pp{}  & \multicolumn{3}{c}{$R=0.2$} & \multicolumn{3}{c}{$R=0.6$}
\tabularnewline \hline & \pTmin & \pTmax & Avg. & \pTmin & \pTmax & Avg.
\tabularnewline \hline \quad Tracking efficiency & 5.9 & 9.1 & 7.7 & 9.4 & 8.9 & 9.0
\tabularnewline \quad Track \pT{} resolution & 1.0 & 1.0 &1.0 & 1.0 & 1.0 & 1.0 
\tabularnewline \quad EMCal nonlinearity & 0.5 & 0.5 & 0.5 & 1.0 & 0.9 & 1.0
\tabularnewline \quad Hadronic correction & 0.2 & 1.2 & 0.5 & 1.2 & 2.1 & 0.9
\tabularnewline \quad Jet matching & 0.1 & 0.0 & 0.0 & 0.2 & 0.2 & 0.2
\tabularnewline \quad PYTHIA fragmentation & 0.5 & 1.0 & 0.4 & 3.1 & 5.6 & 5.8
\tabularnewline \hline \quad Total corr. uncertainty & 6.0 & 9.3 & 7.8 & 10.1 & 10.8 & 10.8

\tabularnewline \hline 
\end{tabular}
\label{table:corruncertaintiesPP}
\end{table}

\begin{table}[!b]
\centering
\caption{Summary of correlated systematic uncertainties on the \PbPb{} jet spectra, for select $R$ and $\pTlead$ thresholds.
The columns \pTmin{} and \pTmax{} are the uncertainties at the minimum and maximum \pTjet{} bin.}
\begin{tabular}{ l ccccccc }
\tabularnewline \hline \quad & \multicolumn{6}{c}{Relative uncertainty (\%)}

\tabularnewline \hline \hline \PbPb{}  & \multicolumn{3}{c}{$R=0.2, 5$ \GeVc} & \multicolumn{3}{c}{$R=0.4, 7$ \GeVc}
\tabularnewline \hline & \pTmin & \pTmax & Avg. & \pTmin & \pTmax & Avg.
\tabularnewline \hline \quad Tracking efficiency & 5.8 & 8.9 & 8.0 & 9.9 & 9.8 & 9.8 
\tabularnewline \quad Track \pT{} resolution  & 1.0 & 1.0 & 1.0 & 1.0 & 1.0 & 1.0
\tabularnewline \quad EMCal nonlinearity & 2.1 & 1.1 & 1.6 & 11.4 & 7.9 & 9.5
\tabularnewline \quad Hadronic correction & 0.8 & 5.9 & 2.0 & 12.8 & 9.9 & 12.4
\tabularnewline \quad Jet matching & 2.0 & 2.0 & 2.0 & 6.0 & 2.0 & 2.8
\tabularnewline \quad PYTHIA fragmentation & 0.8 & 3.6 & 2.0 & 2.8 & 5.1 & 3.8
\tabularnewline \hline \quad Total corr. uncertainty & 6.7 & 11.6 & 9.2 & 20.9 & 16.9 & 19.5

\tabularnewline \hline 
\end{tabular}
\label{table:corruncertaintiesPbPb}
\end{table}

Systematic uncertainties due to the modeling of the EMCal response were included in several ways.
In order to describe the uncertainty in the MC description of the EMCal hadronic response, the subtracted energy in the hadronic correction was varied from
100\% to 70\% of the matched track momentum. 
Moreover, a systematic uncertainty associated with the track-matching criteria was included by changing the \pT{}-dependent track-matching criteria to
\pT{}-independent criteria $\Delta \eta < 0.015,\; \Delta \varphi < 0.03$. 
These two uncertainties were combined in quadrature to form the uncertainty on the EMCal hadronic correction procedure.
In order to describe the uncertainty in the MC description of the EMCal electromagnetic response, in the \pp{} case the photon conversion based non-linearity correction was switched off.
These variations were individually performed both in the RM and the data, and the systematic uncertainty was evaluated by comparing the modified unfolded result to the main result. 
In the \PbPb{} case, there is an additional uncertainty due to the fact that the MC does not exactly describe the cluster energy nonlinearity.
To account for this, different cluster non-linearity corrections are typically applied to data and MC; however, in the \PbPb{} embedding procedure, the 
clusters are mixtures of data and MC cells. The main result was computed by applying the data non-linearity parameterization to the mixed data and MC cells in the embedding procedure.
Therefore, we applied the MC non-linearity parameterization as a systematic variation.
In \PbPb{} collisions for $R=0.4$, the uncertainties on the EMCal non-linearity correction and track matching procedure are large, 
primarily due to unfolding effects, which we do not de-couple in the evaluation of the correlated uncertainties.

We included also a systematic uncertainty associated with the choice of jet matching procedure. For \pp{}, 
the geometrical matching distance was varied from $0.4R$ to $0.8R$ (except for $R=0.1$ from $0.2R$ to $0.9R$), which resulted in an uncertainty of less than 1\% (1.5\%). 
For \PbPb, we varied from a pure geometrical matching to an MC-fraction based approach, 
in which a shared momentum fraction requirement ensures that the matched jet contains more than 50\% of the \pT{} of the MC jet. 
This gave an uncertainty of 2--6\%.

We included also a systematic uncertainty associated with the model-dependent reliance on PYTHIA to unfold the spectra.
In \pp{} collisions, we re-weighted the response matrix according to the jet angularity ($g=\sum_i p_{\mathrm{T},i} r_i / \pTjet$, where $r_i=\sqrt{\Delta \eta^2 + \Delta \varphi^2}$ is the
distance of the $i^{\mathrm{th}}$ constituent from the jet axis) at truth-level. Specifically, we re-weighted the response
matrix such that the 50\% largest angularity jets were weighted an additional $\pm30\%$ relative to the 50\% lowest angularity jets. 
This contributed an uncertainty ranging from $\approx 2\% - 7\%$ depending on the jet $R$, and roughly independent of \pT{}.
The same uncertainties were taken for \PbPb{} collisions.

Tables \ref{table:corruncertaintiesPP} and \ref{table:corruncertaintiesPbPb} illustrate the contributions of the various correlated uncertainties for \pp{} and \PbPb{} collisions.
These uncertainties are expected to be largely independent, so we summed their uncertainties in quadrature.

\subsection{Shape uncertainties}

\begin{table}[!b]
\centering
\caption{Summary of shape systematic uncertainties on the \pp{} jet spectra without a leading track bias, for select $R$.
The columns \pTmin{} and \pTmax{} are the uncertainties at the minimum and maximum \pTjet{} bin.}
\begin{tabular}{ l ccccccc }
\tabularnewline \hline \quad & \multicolumn{6}{c}{Relative uncertainty (\%)}

\tabularnewline \hline \hline \pp{}  & \multicolumn{3}{c}{$R=0.2, 0$ \GeVc} & \multicolumn{3}{c}{$R=0.6, 0$ \GeVc}
\tabularnewline \hline & \pTmin & \pTmax & Avg. & \pTmin & \pTmax & Avg.
\tabularnewline \hline \quad Unfolding method & 0.0 & 16.0 & 3.4 & 2.6 & 16.0 & 4.5
\tabularnewline \quad Reg. parameter & 0.7 & 2.4 & 1.3 & 1.0 & 3.5 & 2.1
\tabularnewline \quad Prior & 1.3 & 0.8 & 0.9 & 0.9 & 3.7 & 2.0
\tabularnewline \quad Input \pT{} range & 0.8 & 3.3 & 1.3 & 0.4 & 3.2 & 1.4
\tabularnewline \hline \quad Total shape uncertainty  & 0.8 & 8.3 & 2.2 & 1.5 & 8.5 & 3.0

\tabularnewline \hline 
\end{tabular}
\label{table:shapeuncertaintiesPP}
\end{table}

\begin{table}[!b]
\centering
\caption{Summary of shape systematic uncertainties on the \PbPb{} jet spectra, for select $R$ and $\pTlead$ thresholds.
The columns \pTmin{} and \pTmax{} are the uncertainties at the minimum and maximum \pTjet{} bin.}
\begin{tabular}{ l ccccccc }
\tabularnewline \hline \quad & \multicolumn{6}{c}{Relative uncertainty (\%)}

\tabularnewline \hline \hline \PbPb{}  & \multicolumn{3}{c}{$R=0.2, 5$ \GeVc} & \multicolumn{3}{c}{$R=0.4, 7$ \GeVc}
\tabularnewline \hline & \pTmin & \pTmax & Avg. & \pTmin & \pTmax & Avg.
\tabularnewline \hline \quad Unfolding method & 7.7 & 10.0 & 5.4 & 30.3 & 2.5 & 18.2
\tabularnewline \quad Reg. parameter & 4.2 & 8.7 & 4.4 & 24.9 & 20.6 & 23.1 
\tabularnewline \quad Prior & 1.5 & 6.7 & 2.4 & 2.3 & 8.3 & 4.2 
\tabularnewline \quad Input \pT{} range & 0.4 & 0.9 & 0.6 & 1.5 & 1.7 & 1.4
\tabularnewline \hline \quad Total shape uncertainty  & 4.4 & 7.4 & 3.8 & 19.6 & 11.2 & 15.5 

\tabularnewline \hline 
\end{tabular}
\label{table:shapeuncertaintiesPbPb}
\end{table}

In order to assign a shape uncertainty arising from the unfolding regularization procedure, we performed several systematic variations:

\begin{itemize}
	\item Variation of the unfolding algorithm: We unfolded with a Bayes-inspired iterative unfolding algorithm \cite{bayes}.
	\item Variation of the regularization parameter: In the SVD unfolding, we varied the regularization parameter $k$ one unit above and below the nominal solution.
	\item Variation of the prior: The SVD algorithm requires a prior distribution as input, which for the main result is the projection of the RM onto 
	the truth axis (before normalization). We varied this input prior either by scaling the main prior by $\pT^{\pm 0.5}$ or replacing it with a jet cross-section produced by POWHEG or the unfolded main result itself.
	\item Variation of the input range: For \PbPb{} (\pp{}) collisions, we varied the measured input range $\pm 5$ ($^{+5}_{-3}$) \GeVc{} around the nominal value for each $R$. 
\end{itemize}

The total shape uncertainty is then the standard deviation of the variations, $\sqrt{\sum_{i=1}^{3} \sigma_{i}^{2} / 4}$, 
where $\sigma_{i}$ is the systematic due to a single variation, since they each comprise independent measurements of the same underlying systematic uncertainty in the regularization.
Tables \ref{table:shapeuncertaintiesPP} and \ref{table:shapeuncertaintiesPbPb} illustrate the contributions of the various shape uncertainties for \pp{} and \PbPb{} collisions.

\subsection{Uncertainties on the jet cross-section ratio}

We computed the correlated systematic uncertainties on the \pp{} jet cross-section ratio by making the same variations as in Section 5.1 on both spectra simultaneously, 
and compared the varied jet cross-section ratio to the main result. This resulted in significant cancellation of the correlated uncertainties between the numerator 
and denominator, as can be seen in Section 6. We computed the shape systematic uncertainties by adding the single spectra shape uncertainties in quadrature. 

It is important to note that the statistical uncertainties of the numerator and denominator are partially correlated, due to error propagation through the unfolding procedure.
We did not, however, take this into account. This may result in a slightly conservative statistical uncertainty estimation, since there may be significant cancellation between 
the two radii. Additionally, we did not use statistically independent samples to form the ratio, 
and so the numerator and denominator are statistically correlated with each other, which may lead
to further slight overestimation of the statistical uncertainties.
 
\section{Results}

\subsection{Inclusive Jet Spectra}

\subsubsection{pp}

\begin{figure}[!t]
\centering{}
\includegraphics[scale=0.8]{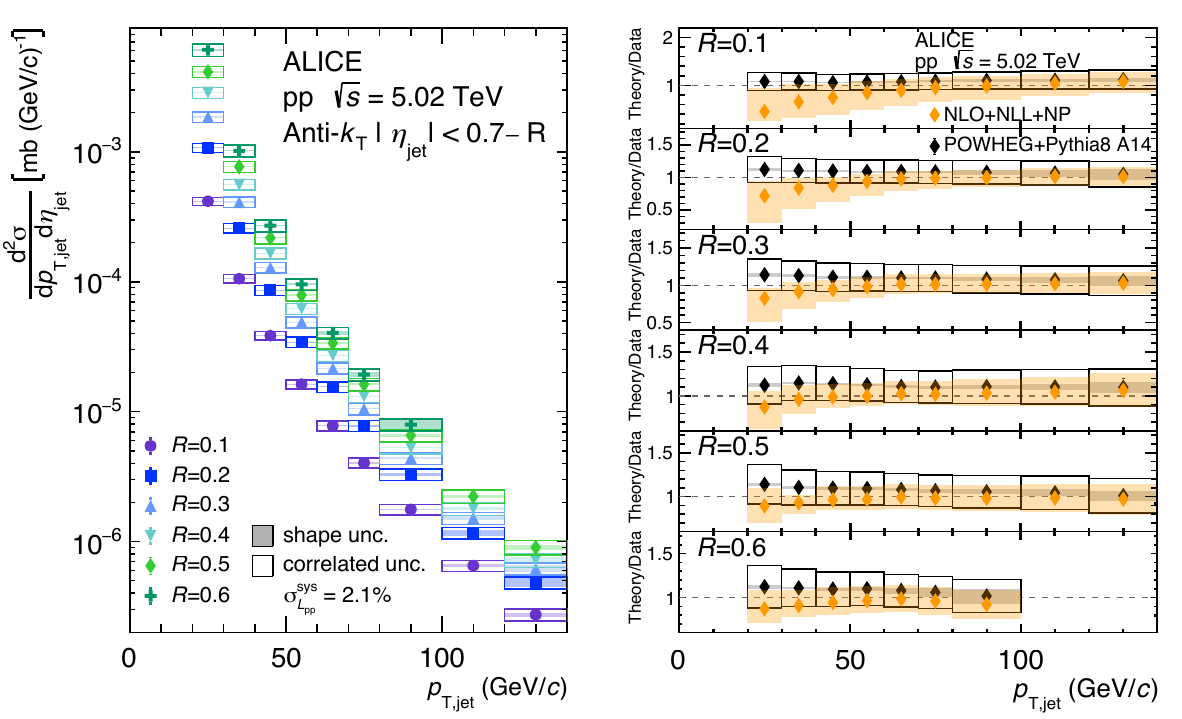}
\caption{Left: Unfolded \pp{} full jet cross-section at $\sqrt{s}=5.02$ TeV for $R=0.1-0.6$. No leading track requirement is imposed.
Right: Ratio of NLO+NLL+NP and POWHEG+PYTHIA8 tune A14 predictions to the measured data. 
The systematic uncertainties in the ratio are denoted by boxes, and are the quadratic sum of the systematic uncertainties in data and the predictions. 
Note that no systematic uncertainties for the non-perturbative correction in the NLO+NLL+NP prediction were included.}
\label{fig:spectraPP}
\end{figure}

We report the \pp{} full jet cross-section for $R=0.1,0.2,0.3,0.4,0.5,0.6$ in Fig. \ref{fig:spectraPP} left.
The cross-sections are reported differentially in \pTjet{} and $\etajet$ as: 
$\frac{\mathrm{d}^{2}\sigma_{jet}}{\mathrm{d}\pTjet \mathrm{d}\etajet} = \frac{1}{\mathcal{L}} \frac{\mathrm{d}^{2}N}{\mathrm{d}\pTjet \mathrm{d}\etajet},$
where we experimentally measured the yield $\frac{\mathrm{d}^{2}N}{\mathrm{d}\pTjet \mathrm{d}\etajet}$ and the integrated luminosity $\mathcal{L}$ \cite{ppXsec}.
The uncertainty on the luminosity is 2.1\%.
The measured jet cross-sections were unfolded for detector and background effects, and are reported at the hadron-level. 
The cross-sections were corrected for the kinematic efficiency and jet reconstruction efficiency, 
as well as the partial azimuthal acceptance of the EMCal and the vertex efficiency.
Note that a leading track requirement was not imposed for the results in Fig. \ref{fig:spectraPP}.

\begin{figure}[!t]
\centering{}
\includegraphics[scale=0.45]{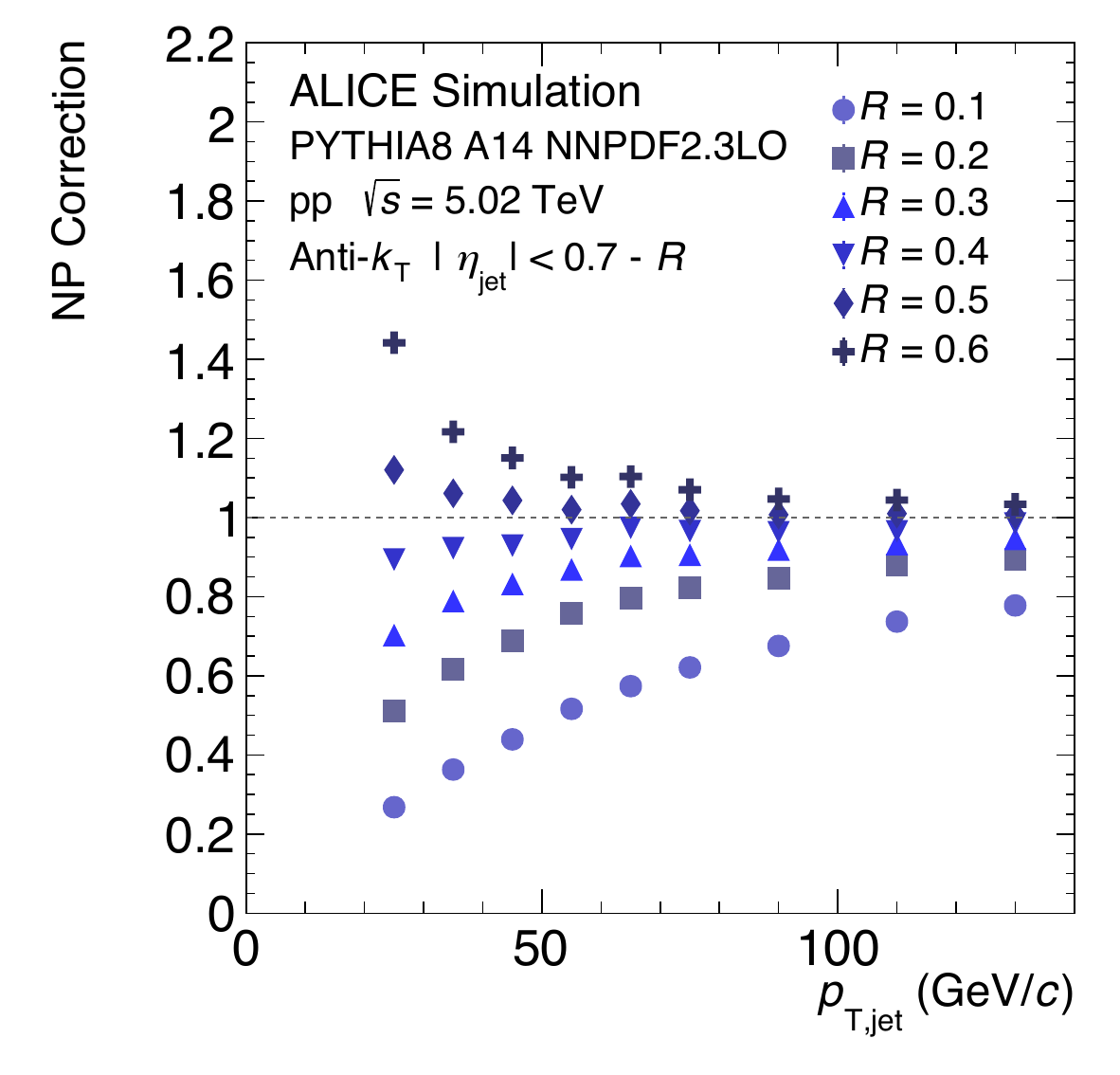}
\caption{Non-perturbative correction factor applied to parton-level NLO+NLL predictions, obtained from PYTHIA8 tune A14 as the ratio of the inclusive jet spectrum
at hadron-level with MPI compared to parton-level without MPI.}
\label{fig:NPCorrection}
\end{figure}

We compare the \pp{} inclusive jet cross-section to two theoretical calculations in Fig. \ref{fig:spectraPP} right.
The predictions denoted NLO+NLL+NP are analytical predictions at NLO with resummation of jet $R$ logarithms and threshold logarithms to NLL accuracy,
performed in a rigorous QCD factorization scheme \cite{SiJF, JointResummation2017, JointResummation2018}. 
The effect of unaccounted higher-order corrections was evaluated by various scale variations, and is included as a systematic uncertainty.
A correction for hadronization and multi-parton interaction (MPI) effects is applied to this prediction, based on PYTHIA8 tune A14,
and is shown in Fig. \ref{fig:NPCorrection}. These non-perturbative (NP) effects become large for low \pTjet{} at both small and large $R$, 
where systematic uncertainties in this correction (beyond the scope of this article) are likely critical.
The predictions use PDF set CT14nlo.
These predictions are seen to be generally consistent with the data, except at low-\pT{} and small-$R$.
This tension may be due to the model-dependent NP correction, which is large in this region.
The experimental data presented in Fig. \ref{fig:spectraPP}, which cover a large range of $R$ down to low \pT{},
and therefore span a wide range of NP effects (from hadronization-dominated at small $R$ to MPI-dominated at large $R$, as seen in Fig. \ref{fig:NPCorrection}),
can be used to further constrain NP effects in \pp{} collisions. This is of relevance both for \pp{} QCD physics
and for interpreting modifications in heavy-ion collisions, which are typically strongest at low \pT{}. 

The predictions denoted POWHEG+PYTHIA8 consist of a MC parton-shower based model using NLO calculations from POWHEG \cite{powheg1} matched to a parton shower and hadronization from PYTHIA8 tune A14.\footnote{The POWHEG reference 
was produced by POWHEG-BOX-V2 at $\sqrt{s} = 5.02$ TeV via the jet pair production process \cite{powheg1, powheg2, powheg3}.
PDF set CT14nlo was used, along with the settings bornktmin$=1$ and bornsuppfact$=70$.
PYTHIA 8.2 tune A14 NNPDF2.3LO was used for the parton shower, which is tuned with ATLAS \pp{} collisions at $\sqrts=7$ TeV using underlying event observables, 
jet substructure observables, and several other observables, not including the inclusive jet cross-section \cite{A14tune}. Merging with PYTHIA was done as in Ref. \cite{powheg4}.
The same set of primary particles was used as described earlier \cite{primaryParticleALICE}.}
Two theoretical uncertainties were computed for these predictions, both in regard to the POWHEG event generation: 
PDF uncertainty, computed as in Ref. \cite{powheg4}, and scale uncertainty, which was computed by varying the 
renormalization and factorization scales. The total theoretical uncertainty on the cross-section was obtained by adding these two contributions in quadrature.
Note that large non-perturbative effects, similar to Fig. \ref{fig:NPCorrection}, are implicitly present in this prediction as well.
The POWHEG+PYTHIA8 predictions are consistent with the measured data for all $R$ and \pTjet{}.
Figure \ref{fig:spectraPP} does not include predictions by PYTHIA alone, since it is well-established that NLO contributions are necessary to obtain the \pp{} inclusive jet cross-section \cite{ppALICE502, NNLO}.

\begin{figure}[!h]
\centering{}
\includegraphics[scale=0.39]{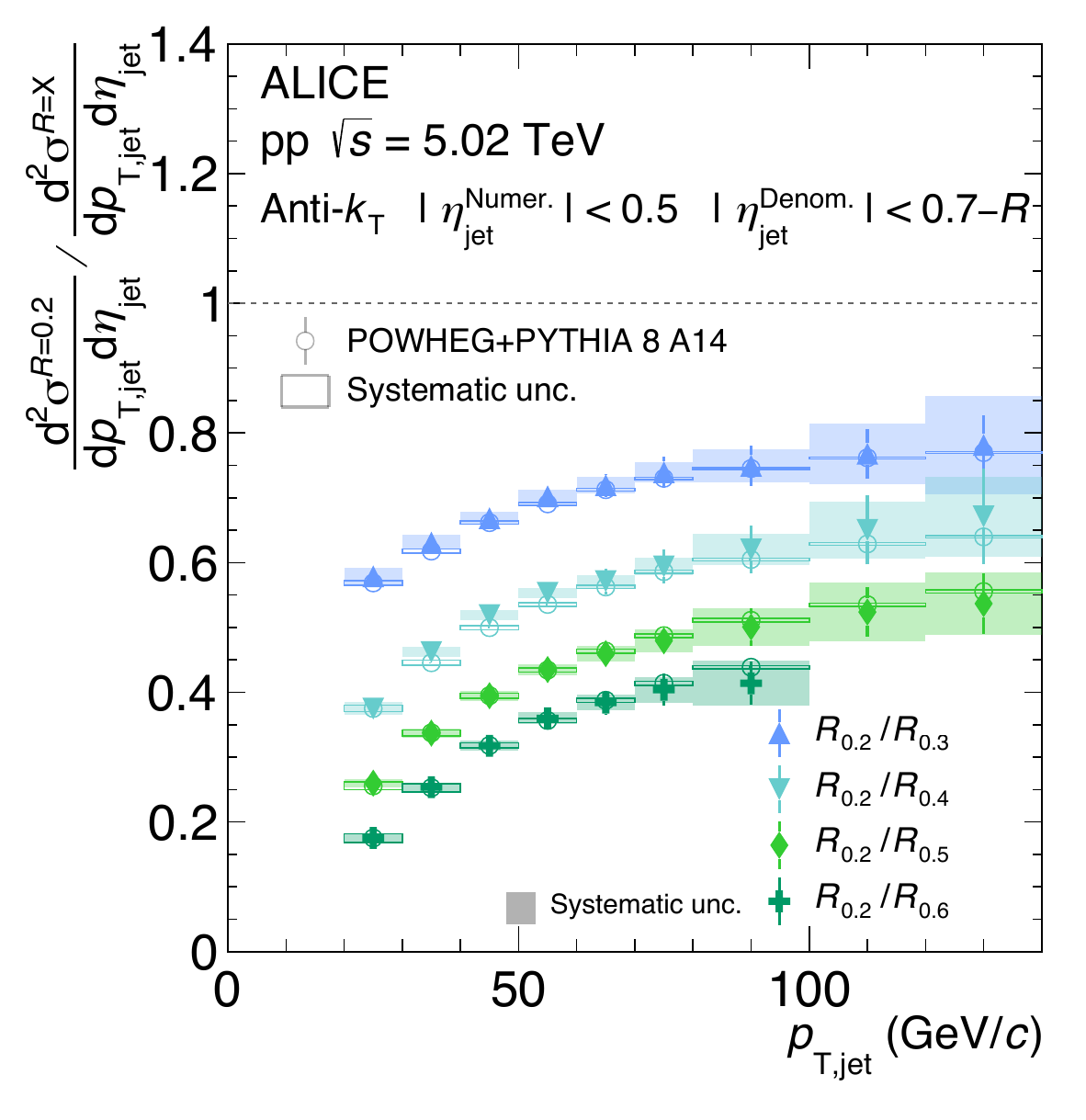}
\includegraphics[scale=0.39]{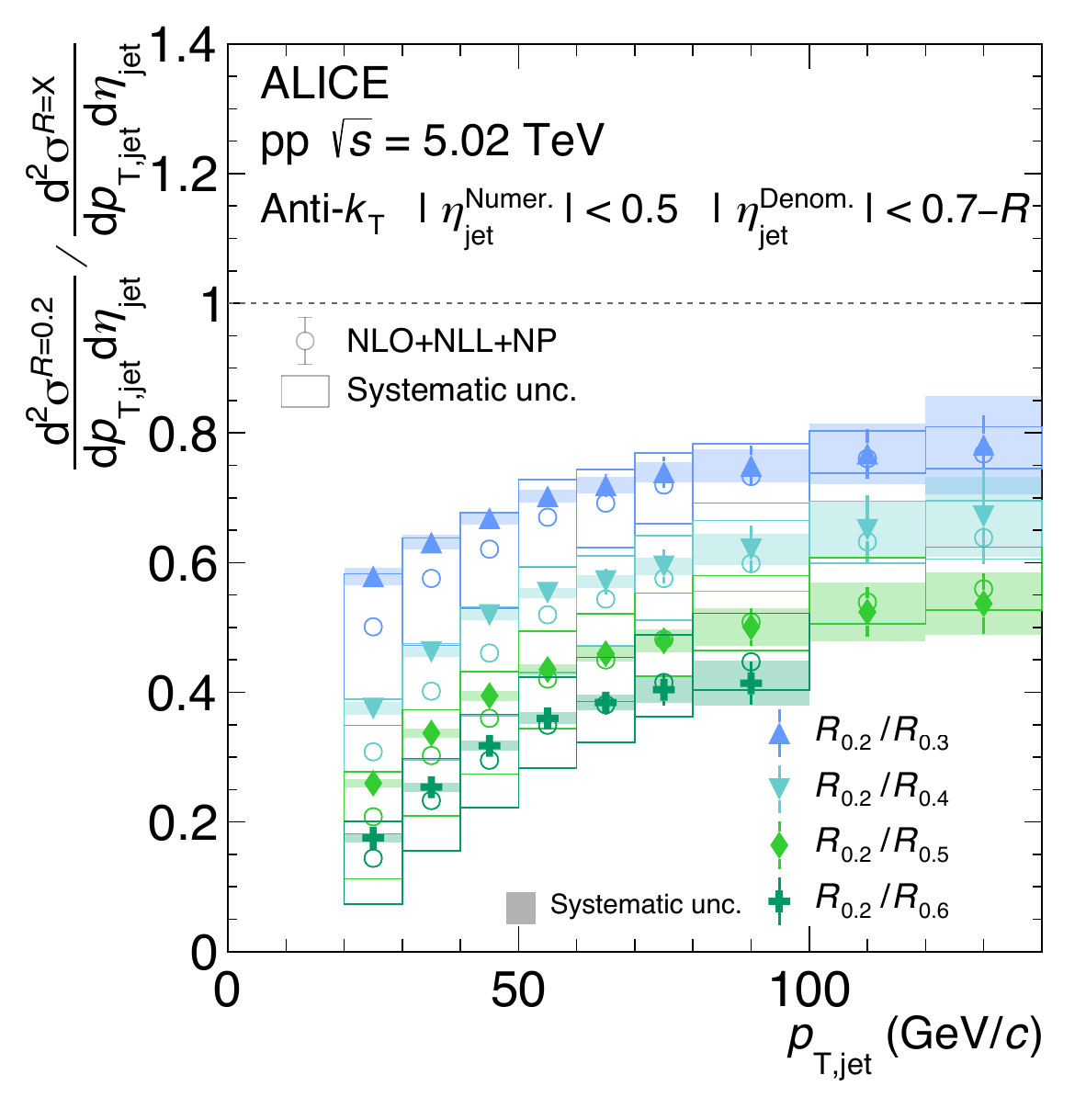}
\includegraphics[scale=0.39]{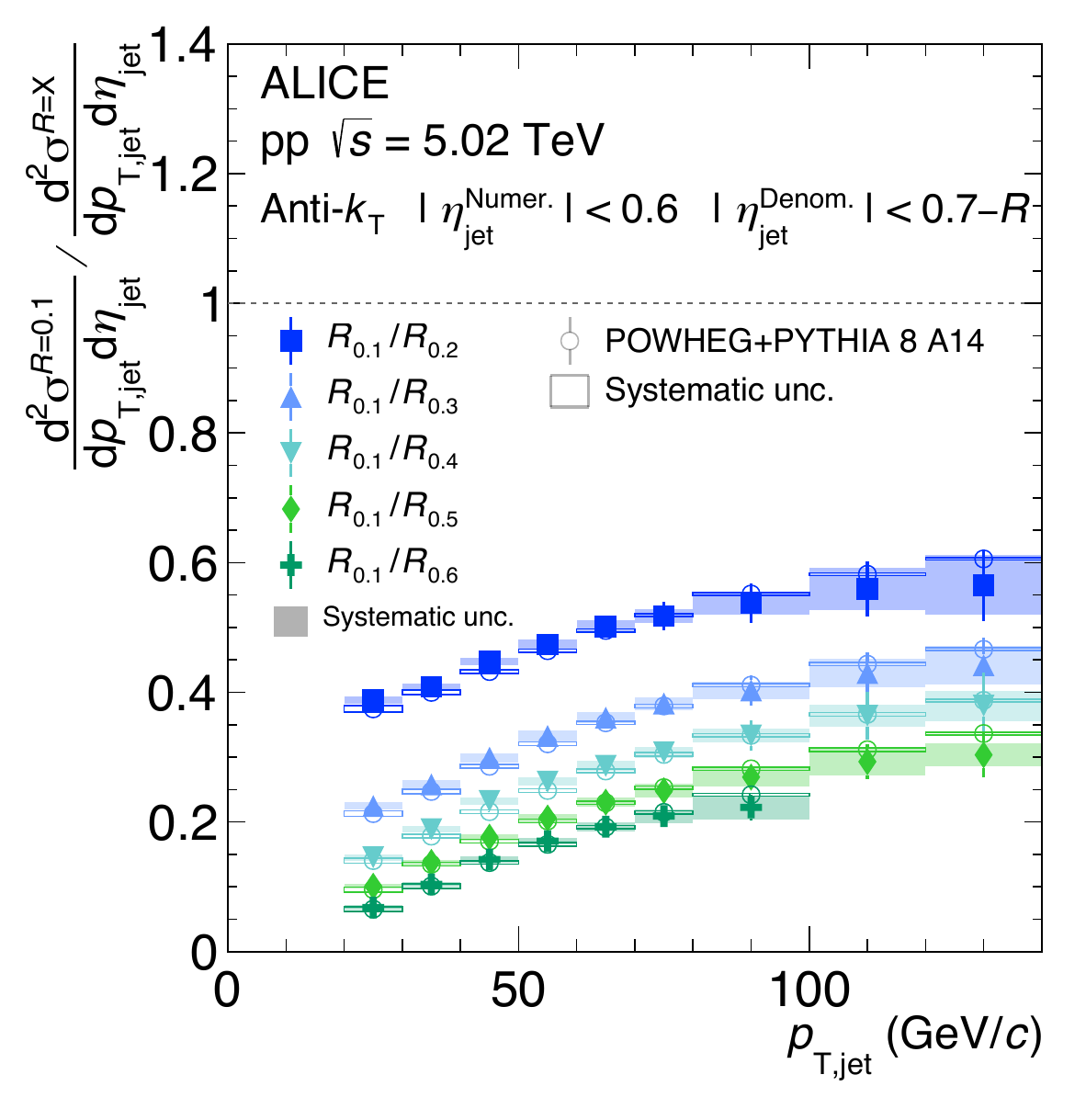}
\includegraphics[scale=0.39]{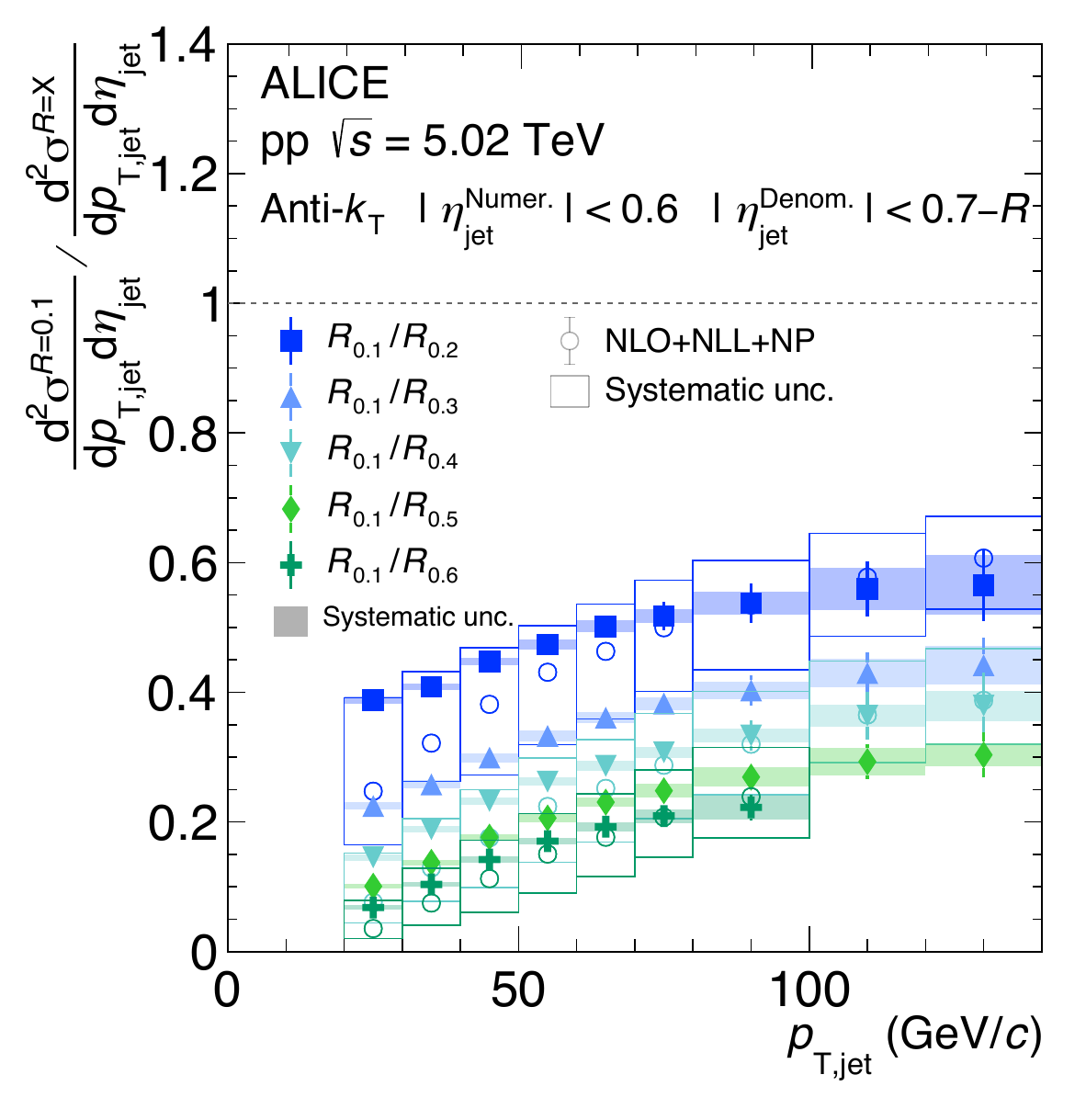}
\caption{Unfolded \pp{} jet cross-section ratios for various $R$. Top panels: ratios of $R=0.2$ to other radii. Bottom panels: ratios of $R=0.1$ to other radii.
The left panels include comparisons to POWHEG + PYTHIA8, and the right panels include comparisons to NLO+NLL+NP.
The experimental correlated systematic uncertainties and shape systematic uncertainties were combined in quadrature into a single systematic uncertainty.
Note that no systematic uncertainties for the non-perturbative correction in the NLO+NLL+NP prediction were included.}
\label{fig:jetCrossSectionRatioPP}
\end{figure}

Figure \ref{fig:jetCrossSectionRatioPP} shows the \pp{} jet cross section ratio for various $R$, built from the spectra in Fig. \ref{fig:spectraPP}.
The top two panels show the ratios of $R=0.2$ to other radii, and the bottom two panels show the ratios of $R=0.1$ to other radii.
The left panels also include comparisons to POWHEG + PYTHIA8, and the right panels include comparisons to NLO+NLL+NP.
Correlated uncertainties largely cancel \cite{ppALICE276, SOYEZ201159}, which allows this observable to elucidate higher-precision effects compared to the inclusive jet cross-section.
The systematic uncertainties on the POWHEG + PYTHIA8 prediction largely cancel as well, and the resulting high-precision comparisons show that the cross-section
ratios are generally well-described by POWHEG + PYTHIA8.
The systematic uncertainties in the NLO+NLL+NP prediction, however, do not substantially cancel, due to the fact that the scale variations include
variation of softer scales which are sensitive to non-perturbative effects; the NLO+NLL+NP predictions are consistent with the measured data within the
size of these large theoretical uncertainties. 

\subsubsection{\PbPb{}}

\begin{figure}[!t]
\centering{}
\includegraphics[scale=0.39]{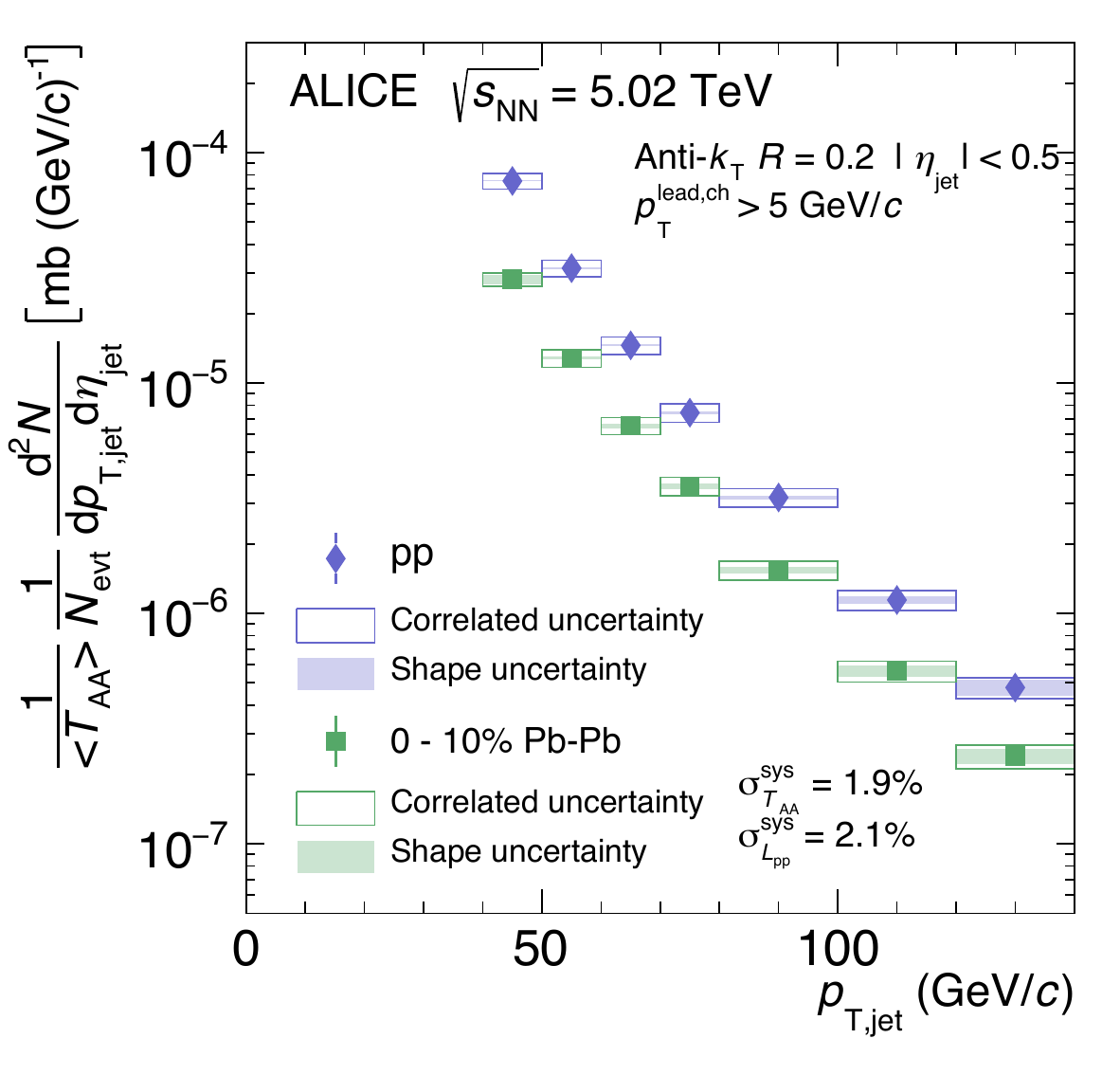}
\includegraphics[scale=0.39]{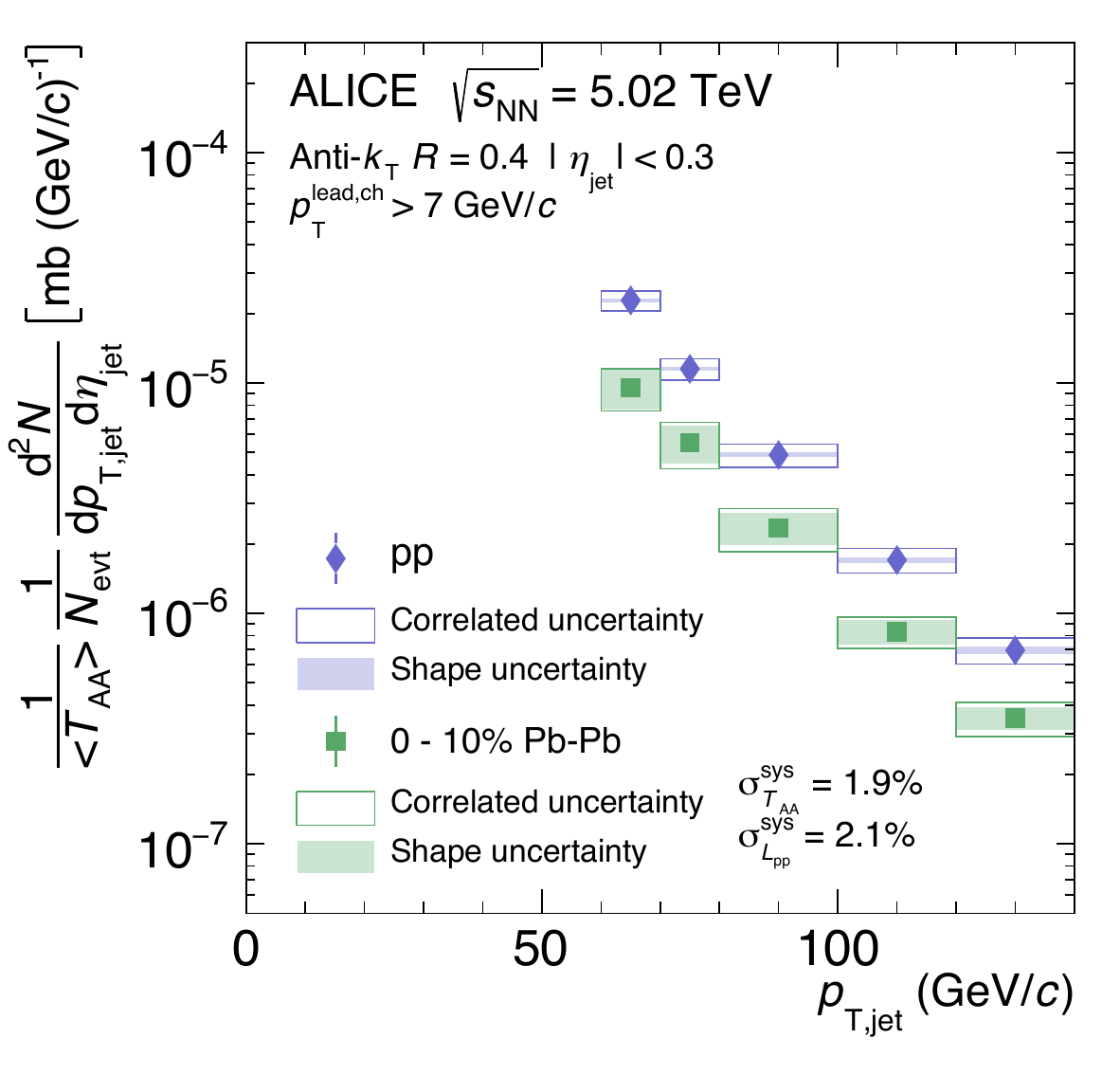}
\caption{Unfolded \pp{} and \PbPb{} full jet spectra at $\sqrts=5.02$ TeV for $R=0.2$ (left), with 5 \GeVc{} leading track requirement, and $R=0.4$ (right), with 7 \GeVc{} leading track requirement.
The \pp{} data points correspond to $\frac{\mathrm{d}^{2}\sigma}{\mathrm{d}\pTjet \mathrm{d}\etajet}$.}
\label{fig:spectra02}
\end{figure}

We report the 0--10\% central \PbPb{} jet spectra for $R=0.2$ and $R=0.4$ in Fig. \ref{fig:spectra02}. 
The spectra are reported differentially in \pTjet{} and $\etajet$ as: $\frac{1}{\Taa} \frac{1}{N_{event}} \frac{\mathrm{d}^{2}N_{jet}^{AA}}{\mathrm{d}\pTjet \mathrm{d}\etajet},$
where $\Taa \equiv \frac{\left<\Ncoll\right>}{\sigma_{inel}^{NN}}$ is the ratio of the number of binary nucleon-nucleon collisions 
to the inelastic nucleon-nucleon cross-section, computed in a Glauber model to be $\Taa = 23.07 \pm 0.44 \; \mathrm{(sys) \; mb^{-1}}$ for 0--10\% centrality.
The jet spectra were unfolded for detector and background effects, and are reported at the hadron-level. 
The spectra were corrected for the kinematic efficiency and jet reconstruction efficiency, as well as the partial azimuthal acceptance of the EMCal.
The $R=0.2$ jets are reported for the range $\pTjet \in [40, 140]$ \GeVc. 
The $R=0.4$ jets are reported for the range $\pTjet \in [60, 140]$ \GeVc. 
The reported intervals were selected based on being insensitive to the combinatorial background, as well as having kinematic efficiency above approximately 80\%.
Note that the reported \pTjet{} intervals extend higher than the measured \pTreco{} range because the kinematic efficiency remains high at larger \pTjet{} due to the JES shift.
A leading track bias of 5 \GeVc{} was required for the $R=0.2$ spectra, while a 7 \GeVc{} bias was required for the $R=0.4$ spectra (both \pp{} reference and \PbPb{})
in order to suppress combinatorial jets in \PbPb{} collisions. We did not attempt to correct to a fully inclusive spectrum, in order to avoid model-dependence.
The \pp{} cross-sections with leading track biases of 5 \GeVc{} and 7 \GeVc{} are plotted alongside the \PbPb{} spectra in Fig. \ref{fig:spectra02}.

\subsection{Jet \Raa}

\begin{figure}[!t]
\centering{}
\includegraphics[scale=0.39]{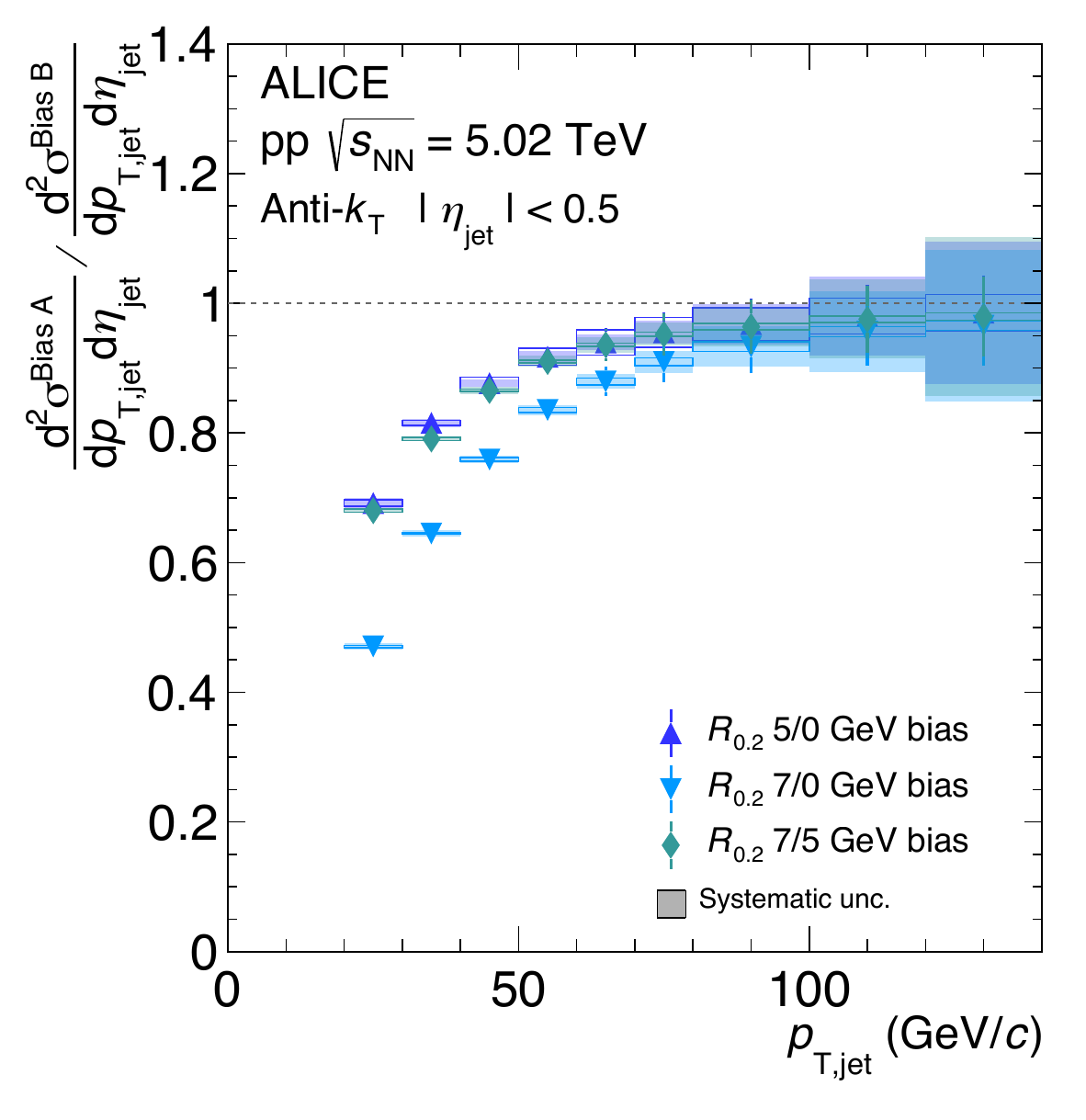}
\includegraphics[scale=0.39]{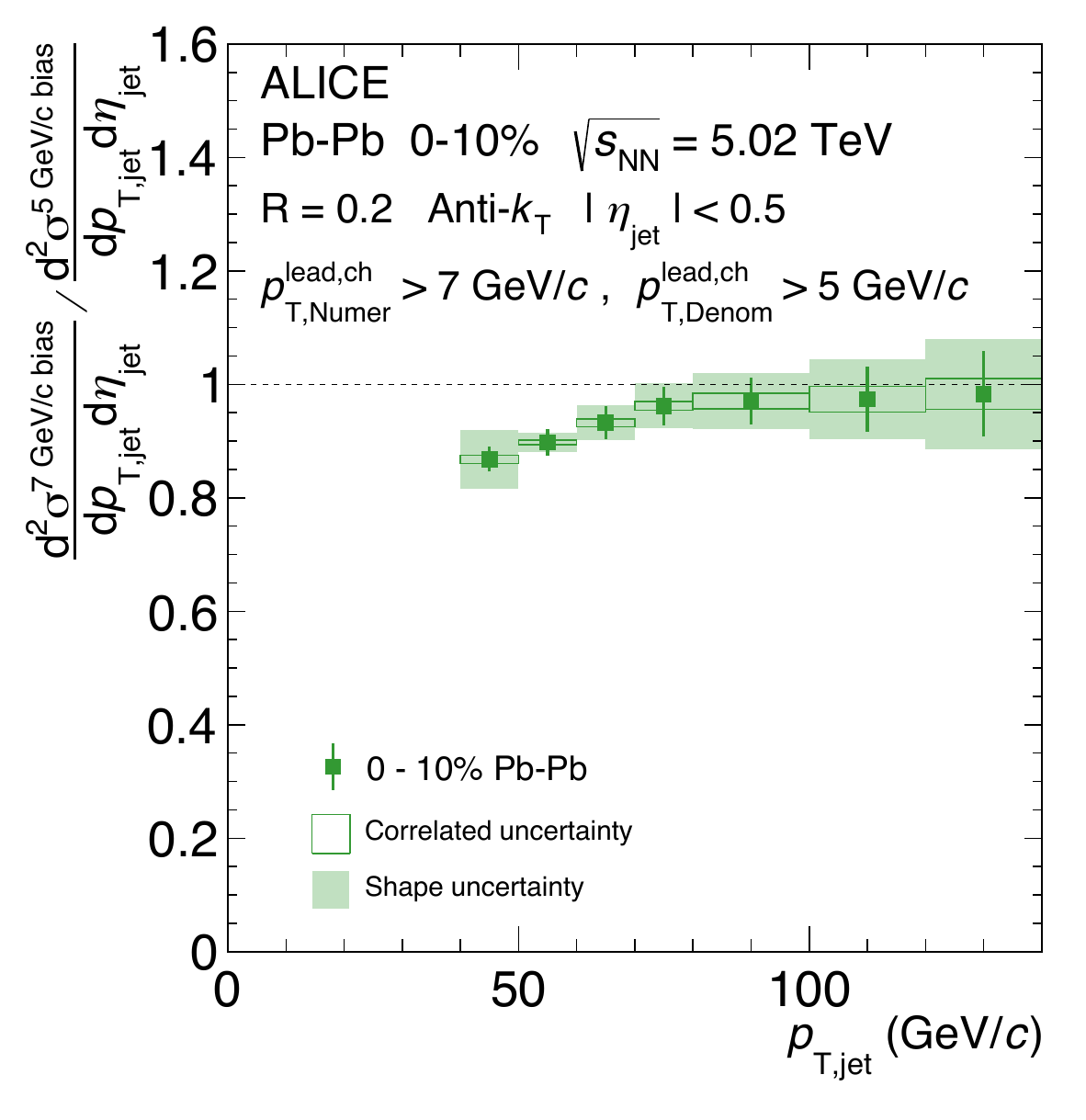}
\caption{Left: Ratio of the \pp{} jet cross-section with various leading charged particle requirements.
Right: Ratio of the $R=0.2$ \PbPb{} jet cross-section with a 7 \GeVc{} leading charged particle requirement compared to a 5 \GeVc{} leading charged particle requirement.}
\label{fig:biasRatio02}
\end{figure}

We report the jet \Raa{} as:

\[
\Raa = \frac { \left. { \frac{1}{N_{\mathrm{event}}} \frac{\mathrm{d}^{2}N}{\mathrm{d}\pTjet \mathrm{d}\etajet}}\right| _{\mathrm{AA}} } {\Taa \left. {\frac{\mathrm{d}^{2}\sigma}{\mathrm{d}\pTjet \mathrm{d}\etajet} }\right| _{\mathrm{pp}} },
\]
namely the ratio of the \PbPb{} and \pp{} spectra plotted above.
While the measured \PbPb{} spectra only report jets satisfying the leading charged hadron requirement, one can choose whether or not to apply the
same requirement for the \pp{} reference, despite that the bias may be different in \pp{} and \PbPb{} collisions. 
To examine the effect of this bias, in Fig. \ref{fig:biasRatio02} we plot the ratio of the $R=0.2$ \pp{} cross-section with either a 0, 5, or 7 \GeVc{}
leading track requirement, as well as the ratio of the $R=0.2$ \PbPb{} jet spectrum with either a 5 or 7 \GeVc{} leading track requirement.
Figure  \ref{fig:biasRatio02} shows that the relative bias between a 5 and 7 \GeVc{} leading track requirement is very similar in \pp{} and \PbPb{} collisions, suggesting that the overall bias in the reported \Raa{} 
may be small compared to the measurement uncertainties. Nevertheless, we report the \Raa{} both with and without a leading track requirement on the \pp{} reference.

\begin{figure}[!t]
\centering{}
\includegraphics[scale=0.39]{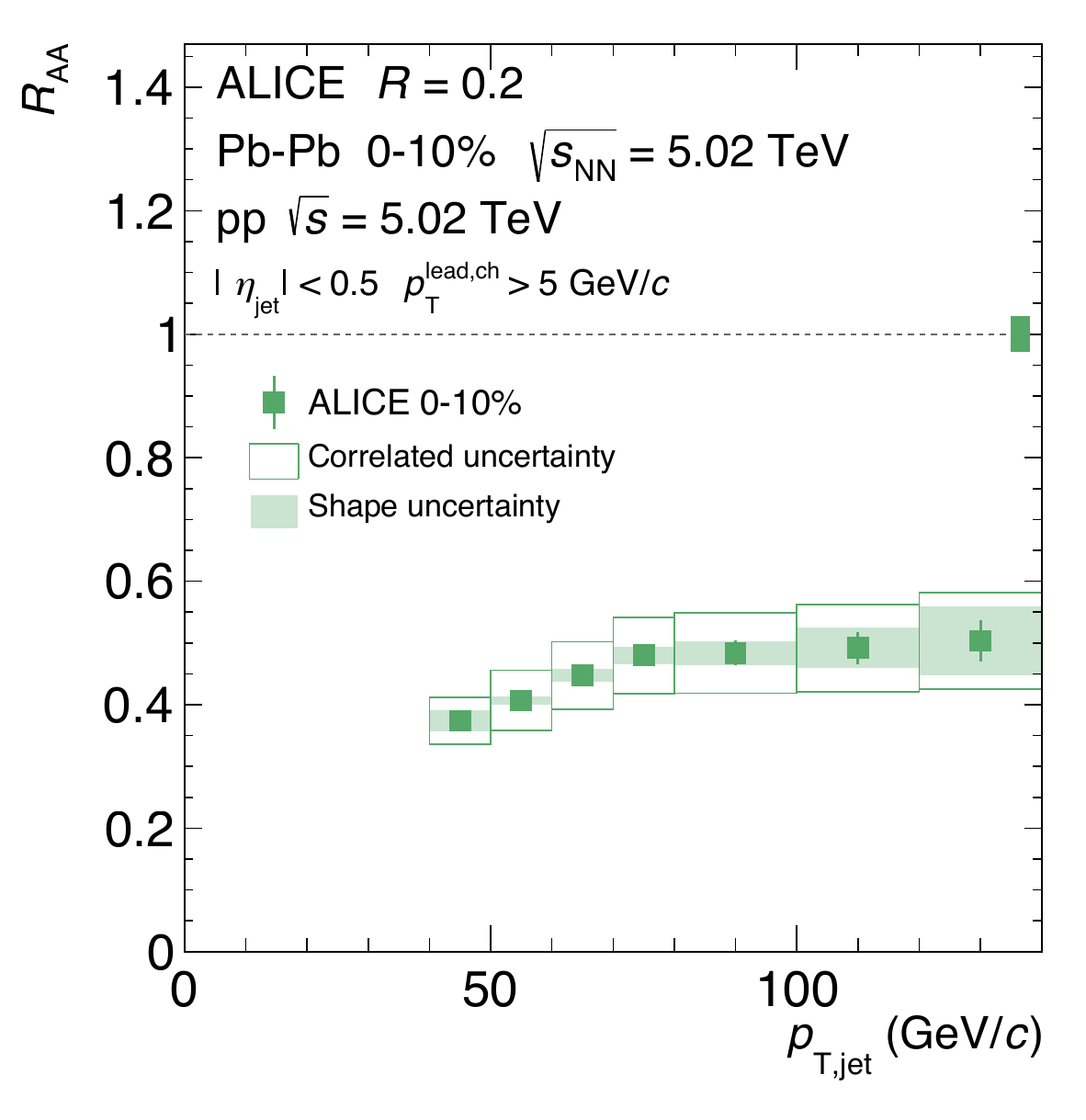}
\includegraphics[scale=0.39]{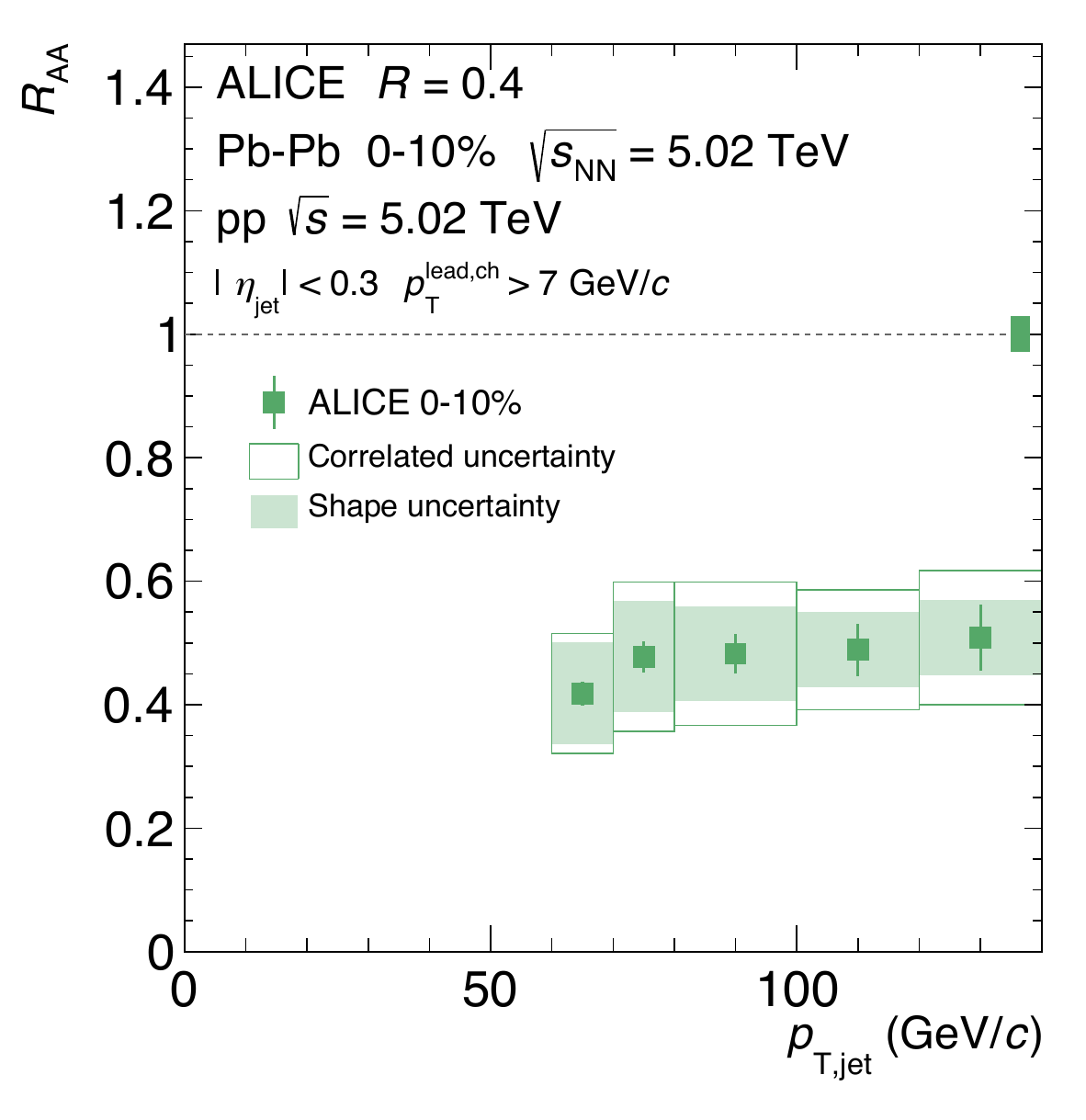}
\includegraphics[scale=0.39]{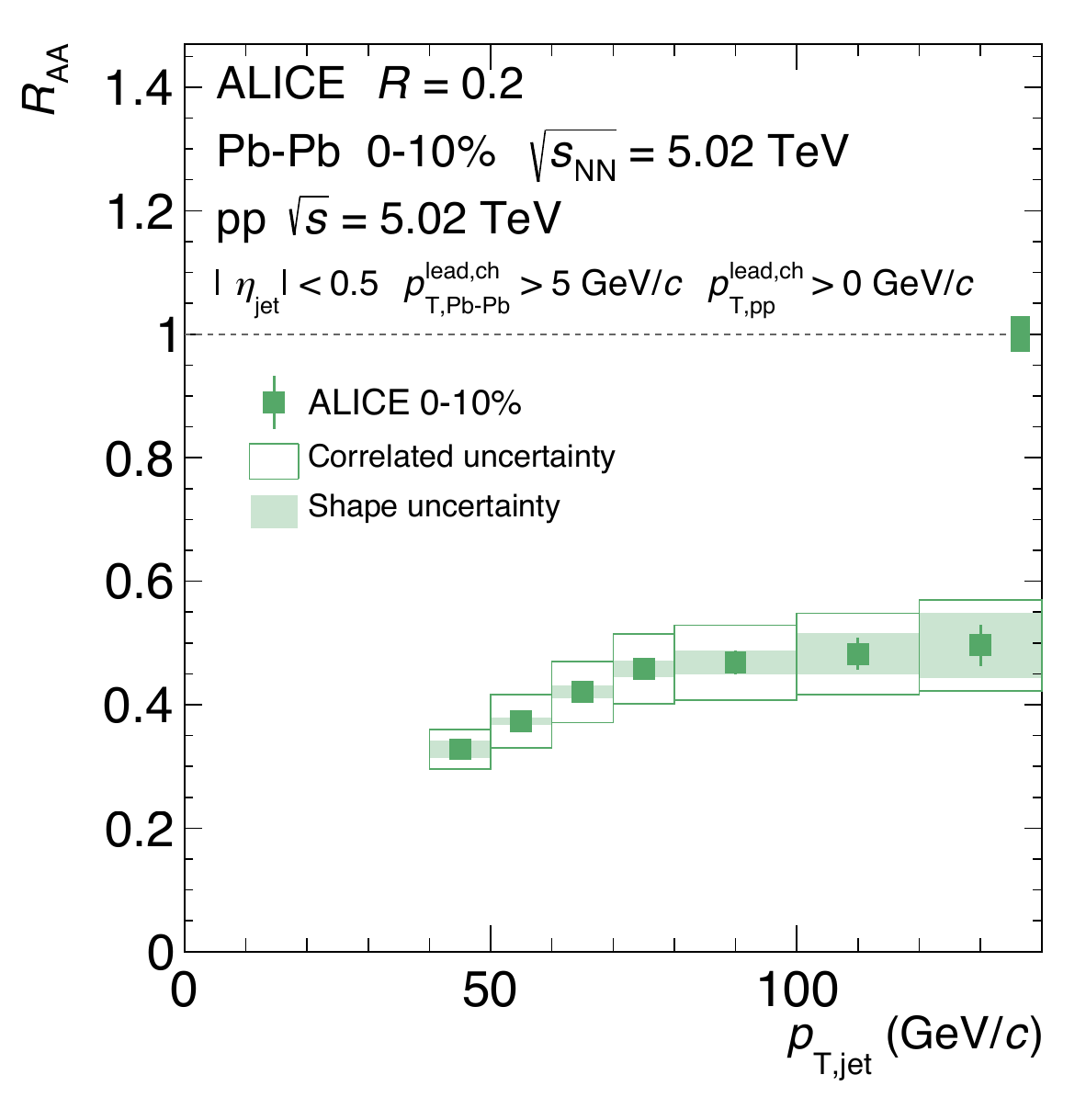}
\includegraphics[scale=0.39]{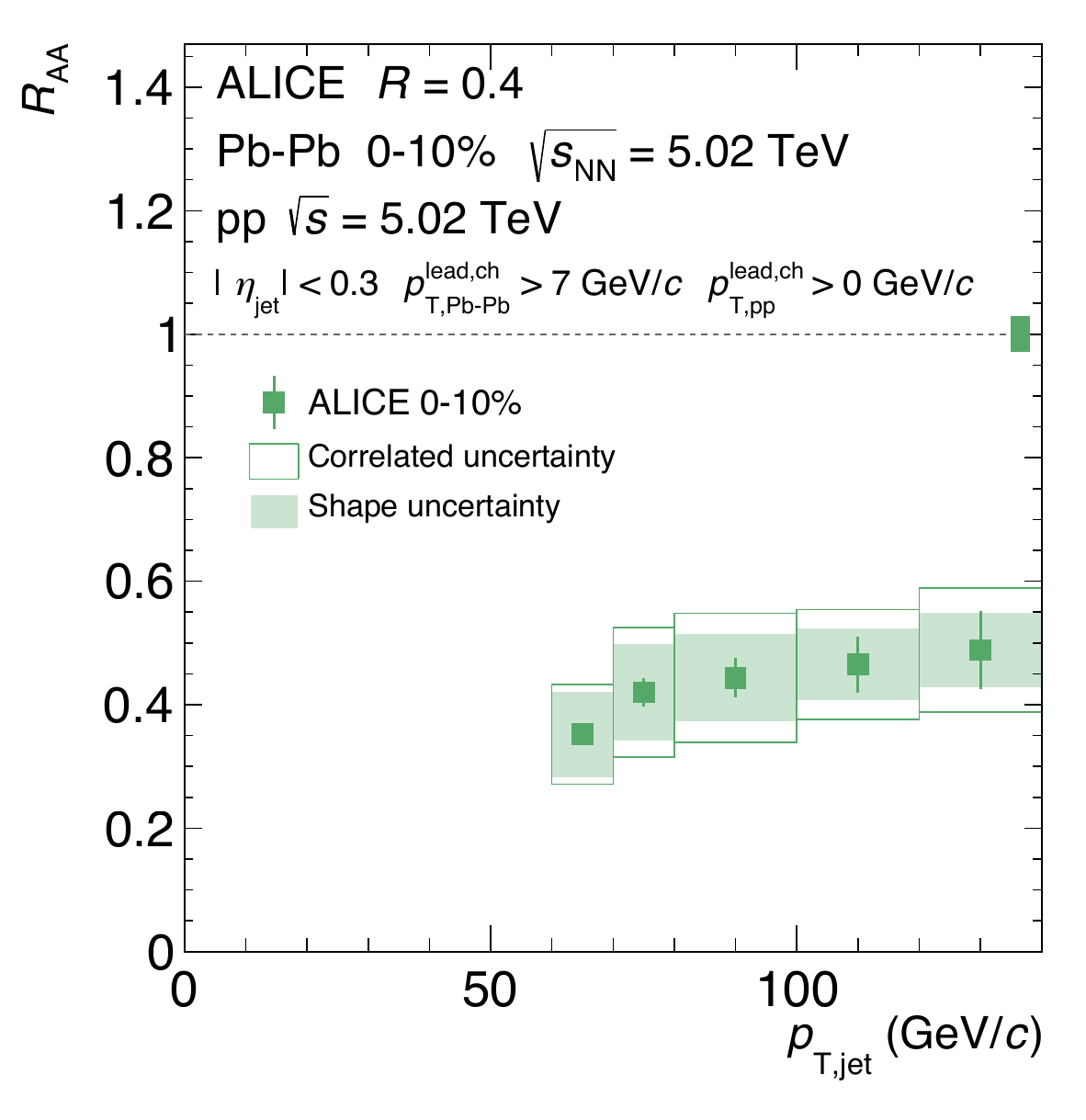}
\caption{Jet \Raa{} at $\sqrts=5.02$ TeV for $R=0.2$ (left) and $R=0.4$ (right). 
In the top panel a leading track requirement is imposed on the \pp{} reference, while in the bottom panel no such requirement is imposed on the \pp{} reference.
The combined \Taa{} uncertainty and \pp{} luminosity uncertainty of 2.8\% is illustrated as a band on the dashed line at $\Raa = 1$.}
\label{fig:Raa}
\end{figure}

Figure \ref{fig:Raa} shows the unfolded full jet \Raa{} for $R=0.2$ and $R=0.4$ jets, both with and without a leading track requirement on the \pp{} reference.
The uncertainties in the \PbPb{} and \pp{} spectra were combined in quadrature.
The jet \Raa{} exhibits strong suppression, and constitutes the first 0--10\% jet \Raa{} measurements at $\sqrts = 5.02$ TeV at low jet \pT{} (i.e. $\pTjet < 100$ \GeVc), and
the first inclusive jet \Raa{} measurements by ALICE extending to $R=0.4$ at any collision energy.
There is visible \pTjet-dependence in the $R=0.2$ case, with stronger suppression at lower \pTjet. 
There is no significant $R$-dependence of the jet \Raa{} within the experimental uncertainties. 
We do not report the jet cross-section ratio for different $R$ in \PbPb{} collisions due to the fact that we found minimal cancellation of uncertainties 
(due to large unfolding uncertainties for $R=0.4$), and it therefore does not convey additional information beyond the reported \Raa{}.

\begin{figure}[!t]
\centering{}
\includegraphics[scale=0.39]{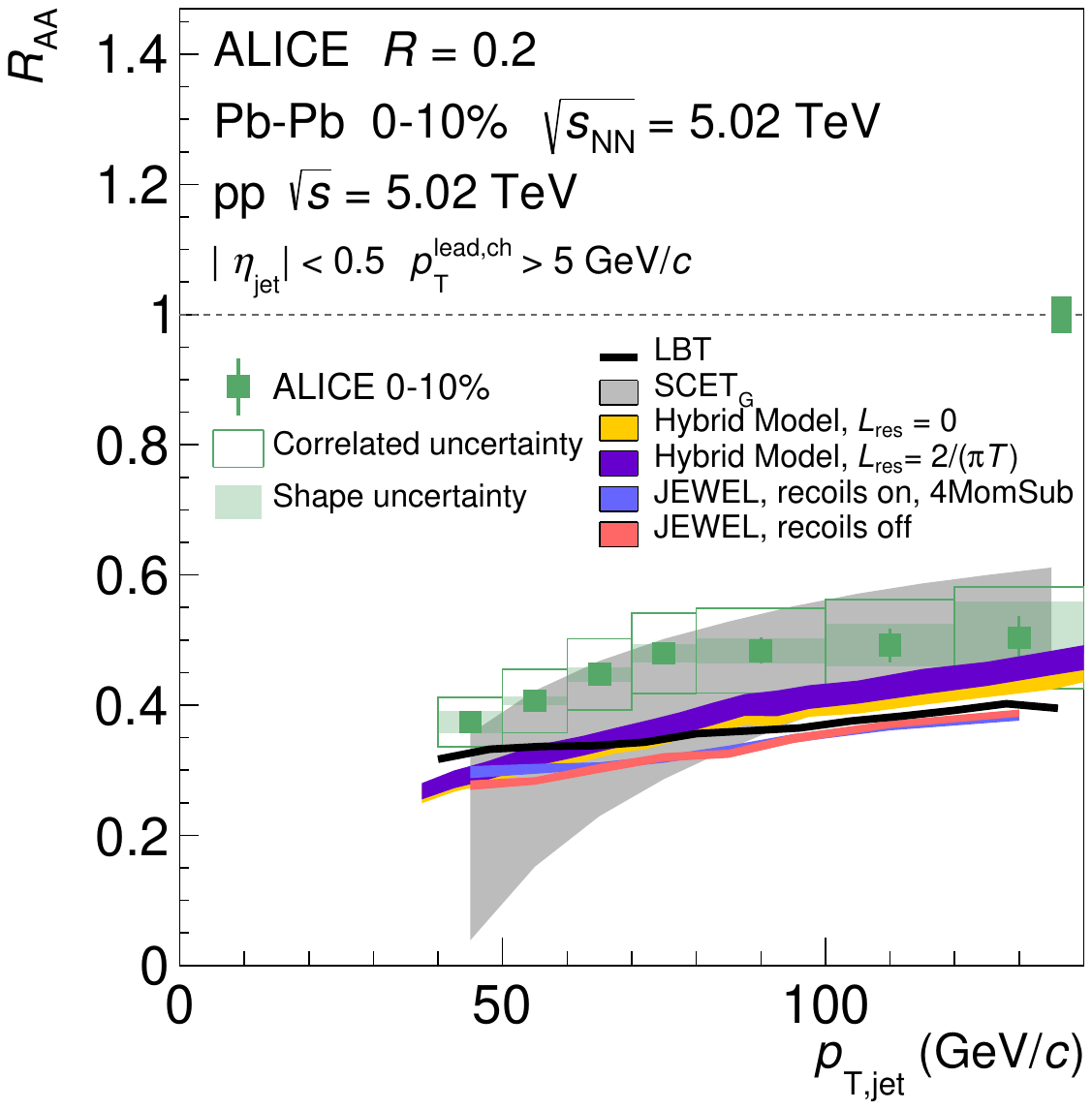}
\includegraphics[scale=0.39]{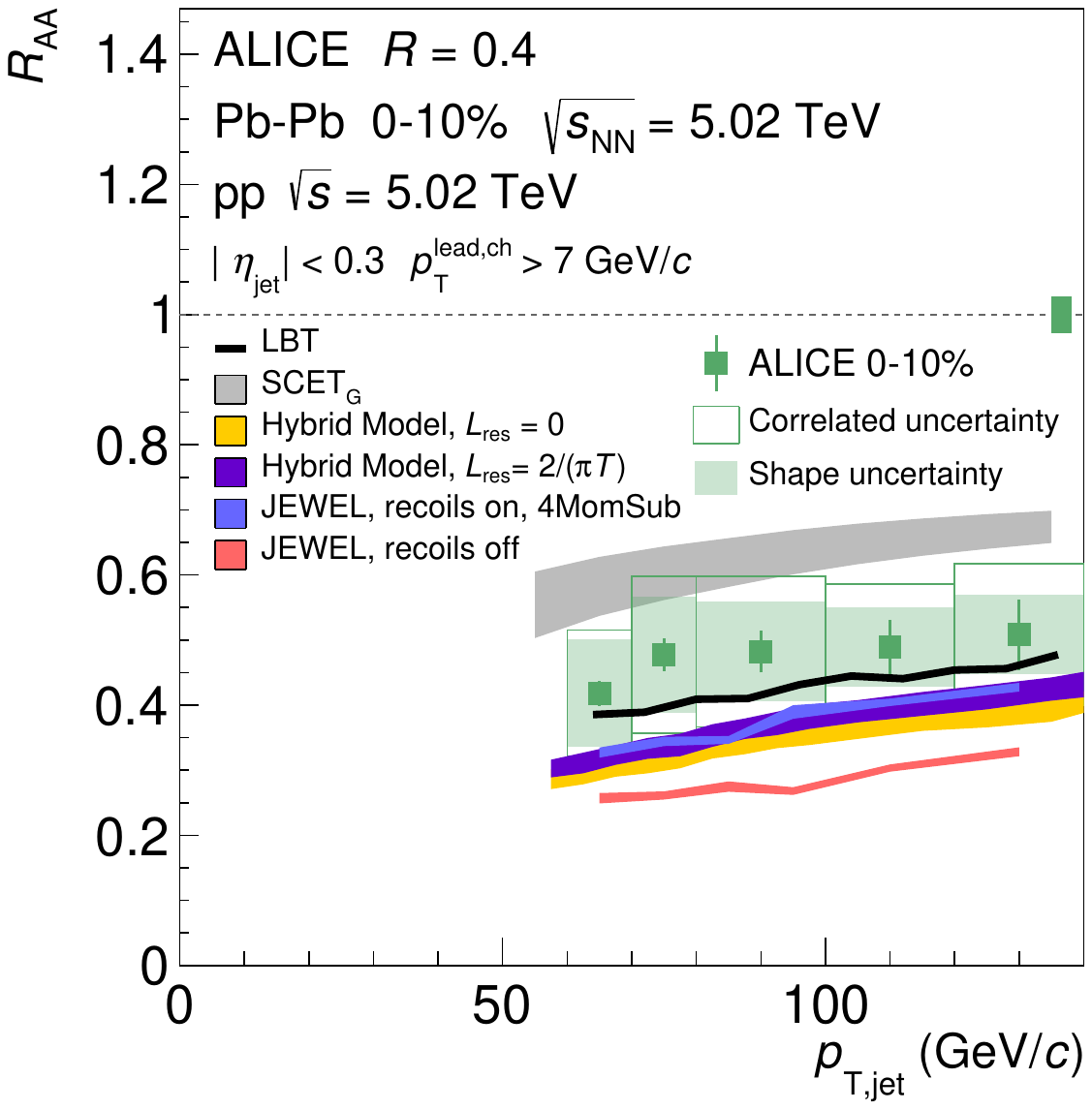}
\caption{Jet \Raa{} at $\sqrts=5.02$ TeV for $R=0.2$ (left) and $R=0.4$ (right) compared to LBT, SCET$_{G}$, Hybrid model, and JEWEL predictions. 
The combined \Taa{} uncertainty and \pp{} luminosity uncertainty of 2.8\% is illustrated as a band on the dashed line at $\Raa = 1$.
Systematic uncertainties are only included for the SCET$_{G}$ and Hybrid model predictions; see text for details.}
\label{fig:RaaTheory}
\end{figure}

We compare these results to four theoretical predictions: the Linear Boltzmann Transport (LBT) model \cite{LBT, LBTconspiracy}, 
Soft Collinear Effective Theory with Glauber gluons (SCET$_{G}$) \cite{SCETG, SCETG2, NLLSCET, SCETsubstructure}, 
the Hybrid model \cite{HybridModel, HybridModelPredictions, HybridModelMediumResponse, HybridModelResolution}, and JEWEL \cite{Jewel, JewelVJet}.
The \Raa{} predictions of these models are compared to the measured data (with the leading track requirements imposed) in Fig. \ref{fig:RaaTheory} for $R=0.2$ and $R=0.4$.
The predictions were all computed using the anti-$k_{\mathrm{T}}$ jet algorithm with $|\etajet|<0.7-R$.
Leading track requirements were only applied by JEWEL (as in data) and the Hybrid model (with 5 \GeVc{} for both radii), for both \pp{} and \PbPb{} collisions.

JEWEL is a Monte Carlo implementation of BDMPS jet energy loss with a parton shower,
and allows the option to include the recoiling thermal medium particles in the jet energy (``recoil on"), or to ignore the recoiling medium particles (``recoil off") \cite{JewelMediumResponse}. 
In the case of including the recoils, the recoil particles free stream and do not interact again with the medium. 
If recoils are included, we perform background subtraction according to the recommended option ``4MomSub".
JEWEL contains several free parameters that are fixed by independent measurements, none of which use high-\pT{} LHC measurements;
we take $T=440$ MeV and $t_{0}=0.4$ fm/c \cite{JewelVJet}.
Note that these predictions do not include systematic uncertainties, but rather only statistical uncertainties. 

The Linear Boltzmann Transport (LBT) model implements pQCD energy loss based on a Higher Twist gluon radiation spectrum induced by elastic scattering, 
and describes the evolution of jet and recoiling medium particles through the thermal medium with linear Boltzmann equations.
An effective strong coupling constant $\alpha_{s}$ is taken as a free parameter fit to experimental data. 
The model calculations are performed according to the methods in Ref. \cite{LBTconspiracy}.
No systematic uncertainties were provided for this calculation.

Soft Collinear Effective Theory with Glauber gluons (SCET$_{G}$) builds on the approach of Soft Collinear Effective Theory (SCET), in which the jet cross-section is factorized into
a ``hard function" corresponding to the initial scattering, and a ``jet function" corresponding to the fragmentation of a hard-scattered parton into a jet. 
In SCET$_{G}$, jet energy loss in heavy-ion collisions is implemented by interactions of jet partons with the hot QCD medium in an effective field theory via the exchange of ``Glauber" gluons, 
encapsulated in an in-medium jet function.
The predictions were performed according to Ref. \cite{NLLSCET} but with minor differences.
The \pp{} jet cross-section was computed to NLO in $\alpha_{s}$, and with a LL resummation in jet $R$.
Medium effects were computed at NLO, but without a resummation in jet $R$ (resulting in large systematic uncertainties for $R=0.2$).
The in-medium splitting functions described above include radiative processes evaluated using 2+1D viscous hydrodynamics, but these predictions do not include collisional energy loss. 
Note that this could have significant impact particularly on the larger radius jets, where it may increase suppression.
The EFT coupling constant between the medium and jets is $g=2.0$.
For \pp{} collisions the CT14nlo PDF was used, and for \PbPb{} collisions, the nCTEQ15FullNuc PDF was used.
Energy loss in cold nuclear matter was also taken into account.
The plotted error band represents the systematic uncertainty obtained by scale variations.

In the Hybrid model, partons are produced by vacuum pQCD, 
and shower according to vacuum pQCD -- but in between these hard splittings, parton energy loss is modeled according to a gauge-gravity 
duality computation in $N=4$ Supersymmetric Yang-Mills at infinitely strong coupling and large $N_{c}$.
Model predictions were provided with two values of $L_{res}$, which describes the scale at which the medium can resolve two split partons.
The medium evolution was modeled by a hydrodynamic expansion.
The plotted error bands represent the combination of statistical and systematic uncertainties.

All models exhibit strong suppression, and produce the same qualitative trend of \Raa{} as a function of \pTjet{}.
In the case $R=0.2$, JEWEL slightly under-predicts the jet \Raa{} regardless of whether medium recoils are included, 
while for $R=0.4$ the ``recoils on" prediction is more consistent with the data.
There is no significant difference between the ``recoil on" or ``recoil off" option in JEWEL for $R=0.2$; 
one expects in general a smaller impact from medium recoil in smaller radius jets.
The LBT model describes the data marginally better, but still shows slight tension. 
Note that the dominant systematic uncertainties in the data are positively correlated between \pTjet{} bins. 
Neither the JEWEL nor LBT predictions include systematic uncertainties.
The SCET$_{G}$ predictions are consistent with the data, 
although the $R=0.2$ prediction has large systematic uncertainties due to a lack of in-medium $\ln R$ re-summation in this calculation.
Additionally, the SCET$_{G}$ calculation did not include collisional energy loss, which may under-estimate suppression for $R=0.4$.
The Hybrid model describes the trend of the data reasonably well, although like the LBT model, exhibits slight tension particularly in the $\pTjet < 100$ \GeVc{} range.
The shapes of the \pTjet{}-dependence differ between the model predictions, most notably between SCET$_{G}$ and the others.
While the experimental uncertainties are larger for $R=0.4$, the model predictions span a wider range of \Raa{} than in the case of $R=0.2$, 
which highlights the importance of measuring the $R$-dependence of the jet \Raa.

\begin{figure}[!t]
\centering{}
\includegraphics[scale=0.39]{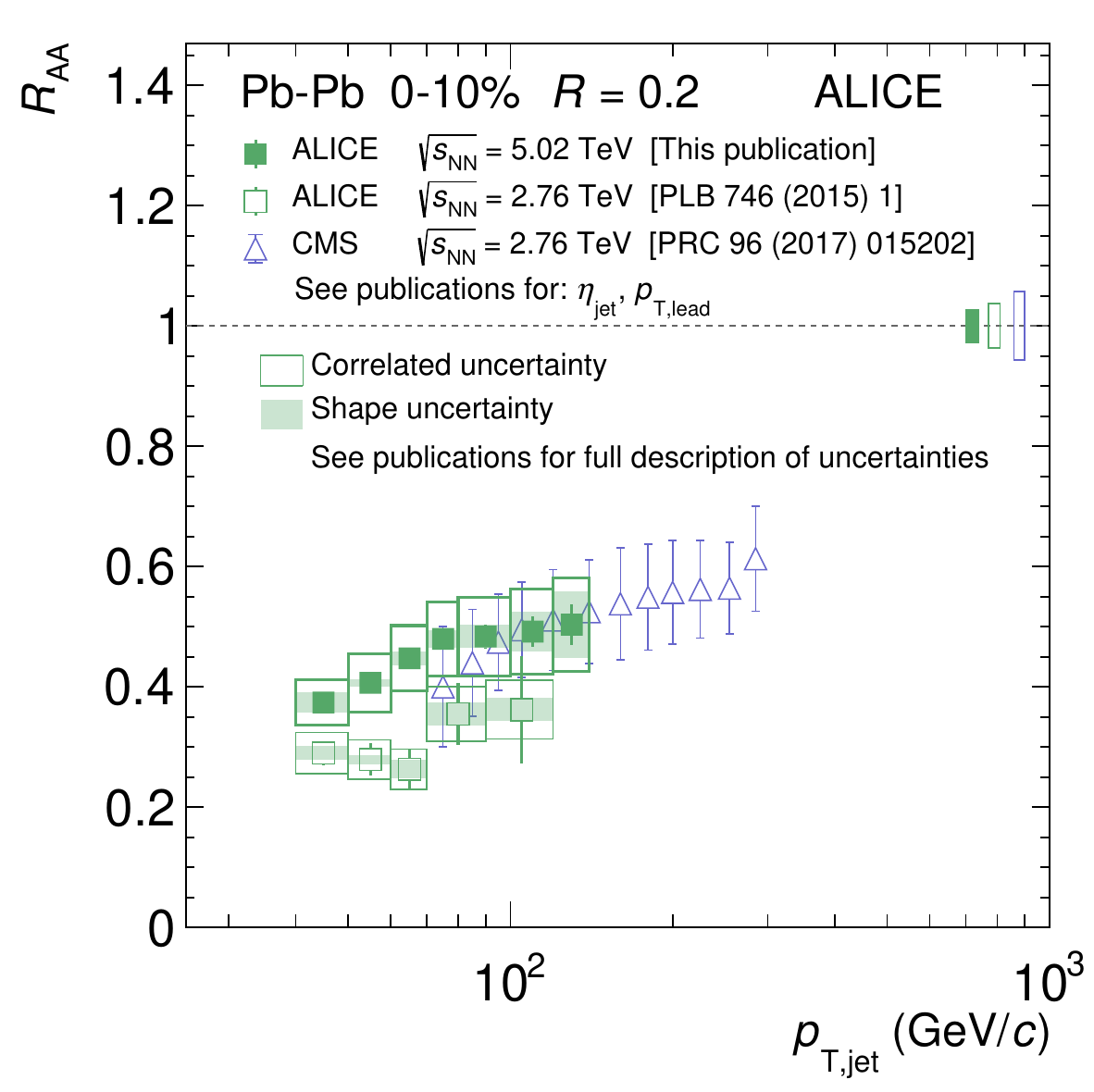}
\includegraphics[scale=0.39]{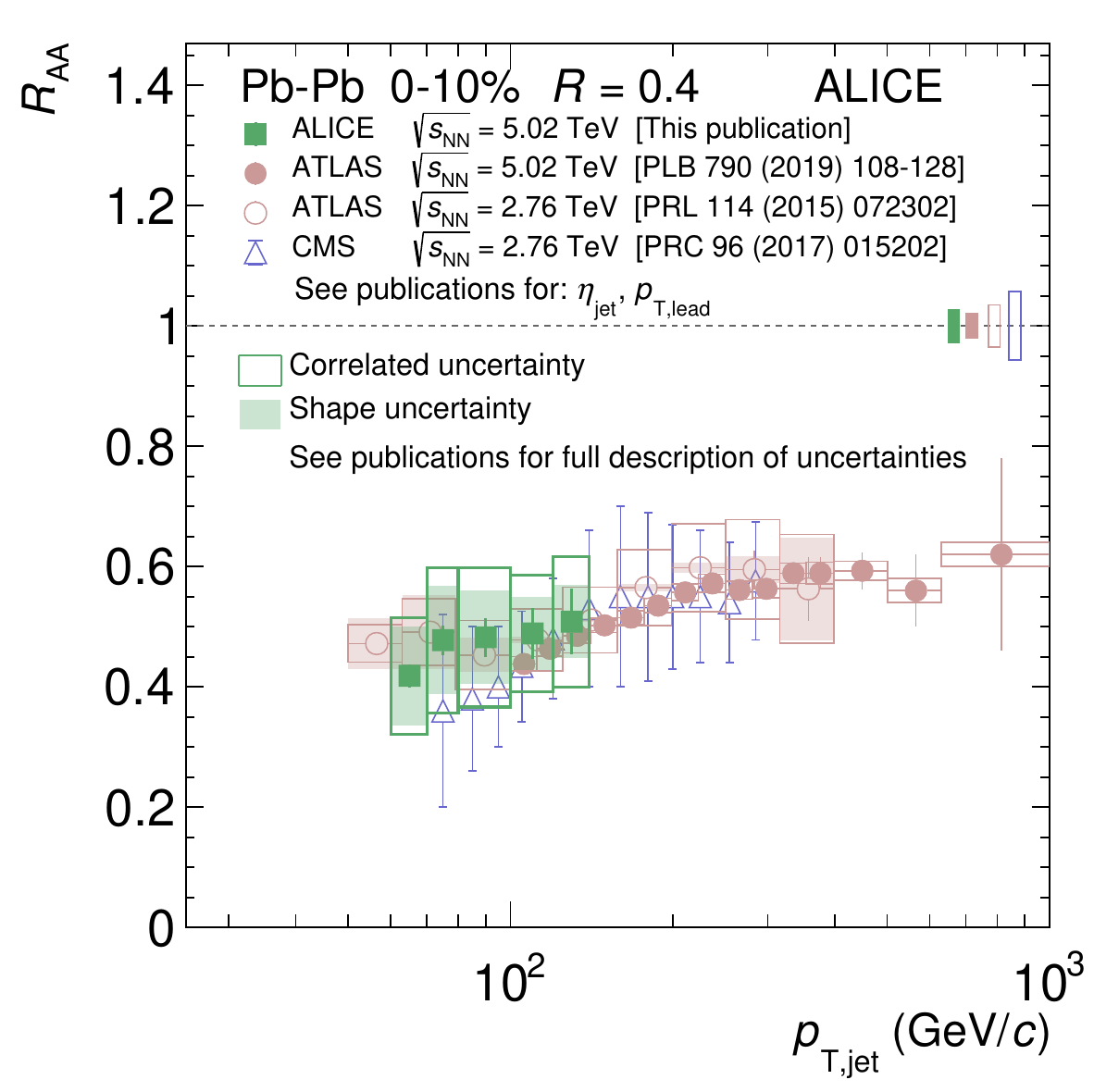}
\caption{Jet \Raa{} in 0-10\% central \PbPb{} collisions for $R=0.2$ (left) and $R=0.4$ (right) for all currently published experimental results. 
Closed markers denote $\sqrts=5.02$ TeV, and open markers denote $\sqrts=2.76$ TeV.}
\label{fig:RaaExp}
\end{figure}

The predictions typically use different strategies for each of the ``non jet energy loss" pieces (initial state, expansion, hadronization, \pp{} reference spectrum),
and do not attempt to incorporate these differences in a systematic uncertainty, which makes a strict quantitative comparison to data difficult.
Moreover, the models fix their free parameters in different ways; JEWEL has not been fit to high-\pTjet{} LHC data, and so it faces the strictest test of all the models presented.
This necessitates investigation of complementary jet observables and global analyses, 
but it also highlights the need to standardize the ingredients of jet energy loss calculations \cite{JetscapeComp}.
The data presented here provide important constraints at low-\pTjet{} on current models as well as for future global analyses.

Figure \ref{fig:RaaExp} shows a comparison of the results in Fig. \ref{fig:Raa} (top) to all currently published experimental results of 0-10\% central jet \Raa{} for $R=0.2$ and $R=0.4$.
This comparison broadly illustrates that there is no clear $R$-dependence or $\sqrt{s}$-dependence of the jet \Raa{} within current experimental precision.
We do not attempt to comment on the comparison of theoretical models with global experimental results, which is beyond the scope of this article.

\section{Conclusion}

We measured the transverse momentum (\pT) spectrum of jets in \pp{} collisions at $\sqrt{s} = 5.02$ TeV and \PbPb{} collisions at $\sqrts = 5.02$ TeV, as well as the jet nuclear modification factor (\Raa),
using charged particles from the tracking system combined with particle information from the electromagnetic calorimeter.
We measured the inclusive jet cross-section in \pp{} collisions for jet resolution parameters $R=0.1-0.6$, which is the largest collection of full jet $R$ measured by ALICE to date. 
We compared these measurements to theoretical predictions at next-to-leading-order (NLO) 
including either a resummation of large logarithms or a matched parton shower. Both predictions describe the data well except with slight tension at low-\pTjet{} for the resummation-based prediction, 
which may be due to either the perturbative calculation or the large non-perturbative corrections at low-\pTjet{}.
We also presented the ratios of jet cross-sections of various $R$, which leverage partial cancellation of systematic uncertainties to obtain high-precision constraints on the $R$-dependence
of the cross-section; the two models considered generally describe these ratios well. 
These data, spanning a large range of $R$ down to low-\pT{}, contain sensitivities to different non-perturbative effects (from hadronization dominated at small $R$ to underlying event dominated at large $R$) and
can be used to constrain the contributions of pQCD, hadronization, and underlying event effects on the inclusive jet cross-section.
These data can further be used to constrain parton distribution functions and the strong coupling constant.

We measured the jet spectrum in \PbPb{} collisions for $R=0.2$ and $R=0.4$, which constitutes the first 0--10\% jet \Raa{} measurements 
at $\sqrts = 5.02$ TeV at $\pTjet < 100$ \GeVc, and the first inclusive jet \Raa{} measurements by ALICE extending to $R=0.4$ at any collision energy.
The measured jet spectrum in \PbPb{} collisions exhibits strong suppression compared to \pp{} collisions, and
for $R=0.2$ the data show stronger suppression at lower \pTjet{} than at higher \pTjet{}.
There is no significant $R$-dependence of the jet \Raa{} within the uncertainties of the measurement. 
Models are able to generally describe the trends of the data, but several models exhibit hints of disagreement with the measurements.
These data provide additional constraints on jet quenching models in heavy-ion collisions, which can be used to extract medium properties such as the transverse momentum diffusion parameter, $\hat{q}$,
as a function of the medium temperature and the jet \pT{}. 
\clearpage
%
%

\newenvironment{acknowledgement}{\relax}{\relax}
\begin{acknowledgement}
\section*{Acknowledgements}

The ALICE Collaboration would like to thank all its engineers and technicians for their invaluable contributions to the construction of the experiment and the CERN accelerator teams for the outstanding performance of the LHC complex.
The ALICE Collaboration gratefully acknowledges the resources and support provided by all Grid centres and the Worldwide LHC Computing Grid (WLCG) collaboration.
The ALICE Collaboration acknowledges the following funding agencies for their support in building and running the ALICE detector:
A. I. Alikhanyan National Science Laboratory (Yerevan Physics Institute) Foundation (ANSL), State Committee of Science and World Federation of Scientists (WFS), Armenia;
Austrian Academy of Sciences, Austrian Science Fund (FWF): [M 2467-N36] and Nationalstiftung f\"{u}r Forschung, Technologie und Entwicklung, Austria;
Ministry of Communications and High Technologies, National Nuclear Research Center, Azerbaijan;
Conselho Nacional de Desenvolvimento Cient\'{\i}fico e Tecnol\'{o}gico (CNPq), Financiadora de Estudos e Projetos (Finep), Funda\c{c}\~{a}o de Amparo \`{a} Pesquisa do Estado de S\~{a}o Paulo (FAPESP) and Universidade Federal do Rio Grande do Sul (UFRGS), Brazil;
Ministry of Education of China (MOEC) , Ministry of Science \& Technology of China (MSTC) and National Natural Science Foundation of China (NSFC), China;
Ministry of Science and Education and Croatian Science Foundation, Croatia;
Centro de Aplicaciones Tecnol\'{o}gicas y Desarrollo Nuclear (CEADEN), Cubaenerg\'{\i}a, Cuba;
Ministry of Education, Youth and Sports of the Czech Republic, Czech Republic;
The Danish Council for Independent Research | Natural Sciences, the VILLUM FONDEN and Danish National Research Foundation (DNRF), Denmark;
Helsinki Institute of Physics (HIP), Finland;
Commissariat \`{a} l'Energie Atomique (CEA), Institut National de Physique Nucl\'{e}aire et de Physique des Particules (IN2P3) and Centre National de la Recherche Scientifique (CNRS) and R\'{e}gion des  Pays de la Loire, France;
Bundesministerium f\"{u}r Bildung und Forschung (BMBF) and GSI Helmholtzzentrum f\"{u}r Schwerionenforschung GmbH, Germany;
General Secretariat for Research and Technology, Ministry of Education, Research and Religions, Greece;
National Research, Development and Innovation Office, Hungary;
Department of Atomic Energy Government of India (DAE), Department of Science and Technology, Government of India (DST), University Grants Commission, Government of India (UGC) and Council of Scientific and Industrial Research (CSIR), India;
Indonesian Institute of Science, Indonesia;
Centro Fermi - Museo Storico della Fisica e Centro Studi e Ricerche Enrico Fermi and Istituto Nazionale di Fisica Nucleare (INFN), Italy;
Institute for Innovative Science and Technology , Nagasaki Institute of Applied Science (IIST), Japanese Ministry of Education, Culture, Sports, Science and Technology (MEXT) and Japan Society for the Promotion of Science (JSPS) KAKENHI, Japan;
Consejo Nacional de Ciencia (CONACYT) y Tecnolog\'{i}a, through Fondo de Cooperaci\'{o}n Internacional en Ciencia y Tecnolog\'{i}a (FONCICYT) and Direcci\'{o}n General de Asuntos del Personal Academico (DGAPA), Mexico;
Nederlandse Organisatie voor Wetenschappelijk Onderzoek (NWO), Netherlands;
The Research Council of Norway, Norway;
Commission on Science and Technology for Sustainable Development in the South (COMSATS), Pakistan;
Pontificia Universidad Cat\'{o}lica del Per\'{u}, Peru;
Ministry of Science and Higher Education and National Science Centre, Poland;
Korea Institute of Science and Technology Information and National Research Foundation of Korea (NRF), Republic of Korea;
Ministry of Education and Scientific Research, Institute of Atomic Physics and Ministry of Research and Innovation and Institute of Atomic Physics, Romania;
Joint Institute for Nuclear Research (JINR), Ministry of Education and Science of the Russian Federation, National Research Centre Kurchatov Institute, Russian Science Foundation and Russian Foundation for Basic Research, Russia;
Ministry of Education, Science, Research and Sport of the Slovak Republic, Slovakia;
National Research Foundation of South Africa, South Africa;
Swedish Research Council (VR) and Knut \& Alice Wallenberg Foundation (KAW), Sweden;
European Organization for Nuclear Research, Switzerland;
Suranaree University of Technology (SUT), National Science and Technology Development Agency (NSDTA) and Office of the Higher Education Commission under NRU project of Thailand, Thailand;
Turkish Atomic Energy Agency (TAEK), Turkey;
National Academy of  Sciences of Ukraine, Ukraine;
Science and Technology Facilities Council (STFC), United Kingdom;
National Science Foundation of the United States of America (NSF) and United States Department of Energy, Office of Nuclear Physics (DOE NP), United States of America.    
We gratefully acknowledge Hai Tao Li, Xiaohui Liu, Daniel Pablos, Felix Ringer, Ivan Vitev, and Xin-Nian Wang for providing theoretical predictions.
\end{acknowledgement}

\bibliography{ALICE_RAA_paper}

\newpage
\appendix
\section{The ALICE Collaboration}
\label{app:collab}

\begingroup
\small
\begin{flushleft}
S.~Acharya\Irefn{org141}\And 
D.~Adamov\'{a}\Irefn{org93}\And 
A.~Adler\Irefn{org73}\And 
J.~Adolfsson\Irefn{org79}\And 
M.M.~Aggarwal\Irefn{org98}\And 
G.~Aglieri Rinella\Irefn{org34}\And 
M.~Agnello\Irefn{org31}\And 
N.~Agrawal\Irefn{org10}\textsuperscript{,}\Irefn{org53}\And 
Z.~Ahammed\Irefn{org141}\And 
S.~Ahmad\Irefn{org17}
% ...
\And
\renewcommand\labelenumi{\textsuperscript{\theenumi}~}

\section*{Affiliation notes}
\renewcommand\theenumi{\roman{enumi}}
\begin{Authlist}
\item \Adef{org*}Deceased
\item \Adef{orgI}Dipartimento DET del Politecnico di Torino, Turin, Italy
\item \Adef{orgII}M.V. Lomonosov Moscow State University, D.V. Skobeltsyn Institute of Nuclear, Physics, Moscow, Russia
\item \Adef{orgIII}Department of Applied Physics, Aligarh Muslim University, Aligarh, India
\item \Adef{orgIV}Institute of Theoretical Physics, University of Wroclaw, Poland
\end{Authlist}

\section*{Collaboration Institutes}
\renewcommand\theenumi{\arabic{enumi}~}
\begin{Authlist}
\item \Idef{org1}A.I. Alikhanyan National Science Laboratory (Yerevan Physics Institute) Foundation, Yerevan, Armenia
\item \Idef{org2}Bogolyubov Institute for Theoretical Physics, National Academy of Sciences of Ukraine, Kiev, Ukraine
\item \Idef{org3}Bose Institute, Department of Physics  and Centre for Astroparticle Physics and Space Science (CAPSS), Kolkata, India
\item \Idef{org4}Budker Institute for Nuclear Physics, Novosibirsk, Russia
\item \Idef{org5}California Polytechnic State University, San Luis Obispo, California, United States
\item \Idef{org6}Central China Normal University, Wuhan, China
\item \Idef{org7}Centre de Calcul de l'IN2P3, Villeurbanne, Lyon, France
\item \Idef{org8}Centro de Aplicaciones Tecnol\'{o}gicas y Desarrollo Nuclear (CEADEN), Havana, Cuba
\item \Idef{org9}Centro de Investigaci\'{o}n y de Estudios Avanzados (CINVESTAV), Mexico City and M\'{e}rida, Mexico
\item \Idef{org10}Centro Fermi - Museo Storico della Fisica e Centro Studi e Ricerche ``Enrico Fermi', Rome, Italy
\item \Idef{org11}Chicago State University, Chicago, Illinois, United States
\item \Idef{org12}China Institute of Atomic Energy, Beijing, China
\item \Idef{org13}Chonbuk National University, Jeonju, Republic of Korea
\item \Idef{org14}Comenius University Bratislava, Faculty of Mathematics, Physics and Informatics, Bratislava, Slovakia
\item \Idef{org15}COMSATS University Islamabad, Islamabad, Pakistan
\item \Idef{org16}Creighton University, Omaha, Nebraska, United States
\item \Idef{org17}Department of Physics, Aligarh Muslim University, Aligarh, India
\item \Idef{org18}Department of Physics, Pusan National University, Pusan, Republic of Korea
\item \Idef{org19}Department of Physics, Sejong University, Seoul, Republic of Korea
\item \Idef{org20}Department of Physics, University of California, Berkeley, California, United States
\item \Idef{org21}Department of Physics, University of Oslo, Oslo, Norway
\item \Idef{org22}Department of Physics and Technology, University of Bergen, Bergen, Norway
\item \Idef{org23}Dipartimento di Fisica dell'Universit\`{a} 'La Sapienza' and Sezione INFN, Rome, Italy
\item \Idef{org24}Dipartimento di Fisica dell'Universit\`{a} and Sezione INFN, Cagliari, Italy
\item \Idef{org25}Dipartimento di Fisica dell'Universit\`{a} and Sezione INFN, Trieste, Italy
\item \Idef{org26}Dipartimento di Fisica dell'Universit\`{a} and Sezione INFN, Turin, Italy
\item \Idef{org27}Dipartimento di Fisica e Astronomia dell'Universit\`{a} and Sezione INFN, Bologna, Italy
\item \Idef{org28}Dipartimento di Fisica e Astronomia dell'Universit\`{a} and Sezione INFN, Catania, Italy
\item \Idef{org29}Dipartimento di Fisica e Astronomia dell'Universit\`{a} and Sezione INFN, Padova, Italy
\item \Idef{org30}Dipartimento di Fisica `E.R.~Caianiello' dell'Universit\`{a} and Gruppo Collegato INFN, Salerno, Italy
\item \Idef{org31}Dipartimento DISAT del Politecnico and Sezione INFN, Turin, Italy
\item \Idef{org32}Dipartimento di Scienze e Innovazione Tecnologica dell'Universit\`{a} del Piemonte Orientale and INFN Sezione di Torino, Alessandria, Italy
\item \Idef{org33}Dipartimento Interateneo di Fisica `M.~Merlin' and Sezione INFN, Bari, Italy
\item \Idef{org34}European Organization for Nuclear Research (CERN), Geneva, Switzerland
\item \Idef{org35}Faculty of Electrical Engineering, Mechanical Engineering and Naval Architecture, University of Split, Split, Croatia
\item \Idef{org36}Faculty of Engineering and Science, Western Norway University of Applied Sciences, Bergen, Norway
\item \Idef{org37}Faculty of Nuclear Sciences and Physical Engineering, Czech Technical University in Prague, Prague, Czech Republic
\item \Idef{org38}Faculty of Science, P.J.~\v{S}af\'{a}rik University, Ko\v{s}ice, Slovakia
\item \Idef{org39}Frankfurt Institute for Advanced Studies, Johann Wolfgang Goethe-Universit\"{a}t Frankfurt, Frankfurt, Germany
\item \Idef{org40}Gangneung-Wonju National University, Gangneung, Republic of Korea
\item \Idef{org41}Gauhati University, Department of Physics, Guwahati, India
\item \Idef{org42}Helmholtz-Institut f\"{u}r Strahlen- und Kernphysik, Rheinische Friedrich-Wilhelms-Universit\"{a}t Bonn, Bonn, Germany
\item \Idef{org43}Helsinki Institute of Physics (HIP), Helsinki, Finland
\item \Idef{org44}High Energy Physics Group,  Universidad Aut\'{o}noma de Puebla, Puebla, Mexico
\item \Idef{org45}Hiroshima University, Hiroshima, Japan
\item \Idef{org46}Hochschule Worms, Zentrum  f\"{u}r Technologietransfer und Telekommunikation (ZTT), Worms, Germany
\item \Idef{org47}Horia Hulubei National Institute of Physics and Nuclear Engineering, Bucharest, Romania
\item \Idef{org48}Indian Institute of Technology Bombay (IIT), Mumbai, India
\item \Idef{org49}Indian Institute of Technology Indore, Indore, India
\item \Idef{org50}Indonesian Institute of Sciences, Jakarta, Indonesia
\item \Idef{org51}INFN, Laboratori Nazionali di Frascati, Frascati, Italy
\item \Idef{org52}INFN, Sezione di Bari, Bari, Italy
\item \Idef{org53}INFN, Sezione di Bologna, Bologna, Italy
\item \Idef{org54}INFN, Sezione di Cagliari, Cagliari, Italy
\item \Idef{org55}INFN, Sezione di Catania, Catania, Italy
\item \Idef{org56}INFN, Sezione di Padova, Padova, Italy
\item \Idef{org57}INFN, Sezione di Roma, Rome, Italy
\item \Idef{org58}INFN, Sezione di Torino, Turin, Italy
\item \Idef{org59}INFN, Sezione di Trieste, Trieste, Italy
\item \Idef{org60}Inha University, Incheon, Republic of Korea
\item \Idef{org61}Institut de Physique Nucl\'{e}aire d'Orsay (IPNO), Institut National de Physique Nucl\'{e}aire et de Physique des Particules (IN2P3/CNRS), Universit\'{e} de Paris-Sud, Universit\'{e} Paris-Saclay, Orsay, France
\item \Idef{org62}Institute for Nuclear Research, Academy of Sciences, Moscow, Russia
\item \Idef{org63}Institute for Subatomic Physics, Utrecht University/Nikhef, Utrecht, Netherlands
\item \Idef{org64}Institute of Experimental Physics, Slovak Academy of Sciences, Ko\v{s}ice, Slovakia
\item \Idef{org65}Institute of Physics, Homi Bhabha National Institute, Bhubaneswar, India
\item \Idef{org66}Institute of Physics of the Czech Academy of Sciences, Prague, Czech Republic
\item \Idef{org67}Institute of Space Science (ISS), Bucharest, Romania
\item \Idef{org68}Institut f\"{u}r Kernphysik, Johann Wolfgang Goethe-Universit\"{a}t Frankfurt, Frankfurt, Germany
\item \Idef{org69}Instituto de Ciencias Nucleares, Universidad Nacional Aut\'{o}noma de M\'{e}xico, Mexico City, Mexico
\item \Idef{org70}Instituto de F\'{i}sica, Universidade Federal do Rio Grande do Sul (UFRGS), Porto Alegre, Brazil
\item \Idef{org71}Instituto de F\'{\i}sica, Universidad Nacional Aut\'{o}noma de M\'{e}xico, Mexico City, Mexico
\item \Idef{org72}iThemba LABS, National Research Foundation, Somerset West, South Africa
\item \Idef{org73}Johann-Wolfgang-Goethe Universit\"{a}t Frankfurt Institut f\"{u}r Informatik, Fachbereich Informatik und Mathematik, Frankfurt, Germany
\item \Idef{org74}Joint Institute for Nuclear Research (JINR), Dubna, Russia
\item \Idef{org75}Korea Institute of Science and Technology Information, Daejeon, Republic of Korea
\item \Idef{org76}KTO Karatay University, Konya, Turkey
\item \Idef{org77}Laboratoire de Physique Subatomique et de Cosmologie, Universit\'{e} Grenoble-Alpes, CNRS-IN2P3, Grenoble, France
\item \Idef{org78}Lawrence Berkeley National Laboratory, Berkeley, California, United States
\item \Idef{org79}Lund University Department of Physics, Division of Particle Physics, Lund, Sweden
\item \Idef{org80}Nagasaki Institute of Applied Science, Nagasaki, Japan
\item \Idef{org81}Nara Women{'}s University (NWU), Nara, Japan
\item \Idef{org82}National and Kapodistrian University of Athens, School of Science, Department of Physics , Athens, Greece
\item \Idef{org83}National Centre for Nuclear Research, Warsaw, Poland
\item \Idef{org84}National Institute of Science Education and Research, Homi Bhabha National Institute, Jatni, India
\item \Idef{org85}National Nuclear Research Center, Baku, Azerbaijan
\item \Idef{org86}National Research Centre Kurchatov Institute, Moscow, Russia
\item \Idef{org87}Niels Bohr Institute, University of Copenhagen, Copenhagen, Denmark
\item \Idef{org88}Nikhef, National institute for subatomic physics, Amsterdam, Netherlands
\item \Idef{org89}NRC Kurchatov Institute IHEP, Protvino, Russia
\item \Idef{org90}NRC «Kurchatov Institute»  - ITEP, Moscow, Russia
\item \Idef{org91}NRNU Moscow Engineering Physics Institute, Moscow, Russia
\item \Idef{org92}Nuclear Physics Group, STFC Daresbury Laboratory, Daresbury, United Kingdom
\item \Idef{org93}Nuclear Physics Institute of the Czech Academy of Sciences, \v{R}e\v{z} u Prahy, Czech Republic
\item \Idef{org94}Oak Ridge National Laboratory, Oak Ridge, Tennessee, United States
\item \Idef{org95}Ohio State University, Columbus, Ohio, United States
\item \Idef{org96}Petersburg Nuclear Physics Institute, Gatchina, Russia
\item \Idef{org97}Physics department, Faculty of science, University of Zagreb, Zagreb, Croatia
\item \Idef{org98}Physics Department, Panjab University, Chandigarh, India
\item \Idef{org99}Physics Department, University of Jammu, Jammu, India
\item \Idef{org100}Physics Department, University of Rajasthan, Jaipur, India
\item \Idef{org101}Physikalisches Institut, Eberhard-Karls-Universit\"{a}t T\"{u}bingen, T\"{u}bingen, Germany
\item \Idef{org102}Physikalisches Institut, Ruprecht-Karls-Universit\"{a}t Heidelberg, Heidelberg, Germany
\item \Idef{org103}Physik Department, Technische Universit\"{a}t M\"{u}nchen, Munich, Germany
\item \Idef{org104}Politecnico di Bari, Bari, Italy
\item \Idef{org105}Research Division and ExtreMe Matter Institute EMMI, GSI Helmholtzzentrum f\"ur Schwerionenforschung GmbH, Darmstadt, Germany
\item \Idef{org106}Rudjer Bo\v{s}kovi\'{c} Institute, Zagreb, Croatia
\item \Idef{org107}Russian Federal Nuclear Center (VNIIEF), Sarov, Russia
\item \Idef{org108}Saha Institute of Nuclear Physics, Homi Bhabha National Institute, Kolkata, India
\item \Idef{org109}School of Physics and Astronomy, University of Birmingham, Birmingham, United Kingdom
\item \Idef{org110}Secci\'{o}n F\'{\i}sica, Departamento de Ciencias, Pontificia Universidad Cat\'{o}lica del Per\'{u}, Lima, Peru
\item \Idef{org111}Shanghai Institute of Applied Physics, Shanghai, China
\item \Idef{org112}St. Petersburg State University, St. Petersburg, Russia
\item \Idef{org113}Stefan Meyer Institut f\"{u}r Subatomare Physik (SMI), Vienna, Austria
\item \Idef{org114}SUBATECH, IMT Atlantique, Universit\'{e} de Nantes, CNRS-IN2P3, Nantes, France
\item \Idef{org115}Suranaree University of Technology, Nakhon Ratchasima, Thailand
\item \Idef{org116}Technical University of Ko\v{s}ice, Ko\v{s}ice, Slovakia
\item \Idef{org117}Technische Universit\"{a}t M\"{u}nchen, Excellence Cluster 'Universe', Munich, Germany
\item \Idef{org118}The Henryk Niewodniczanski Institute of Nuclear Physics, Polish Academy of Sciences, Cracow, Poland
\item \Idef{org119}The University of Texas at Austin, Austin, Texas, United States
\item \Idef{org120}Universidad Aut\'{o}noma de Sinaloa, Culiac\'{a}n, Mexico
\item \Idef{org121}Universidade de S\~{a}o Paulo (USP), S\~{a}o Paulo, Brazil
\item \Idef{org122}Universidade Estadual de Campinas (UNICAMP), Campinas, Brazil
\item \Idef{org123}Universidade Federal do ABC, Santo Andre, Brazil
\item \Idef{org124}University of Cape Town, Cape Town, South Africa
\item \Idef{org125}University of Houston, Houston, Texas, United States
\item \Idef{org126}University of Jyv\"{a}skyl\"{a}, Jyv\"{a}skyl\"{a}, Finland
\item \Idef{org127}University of Liverpool, Liverpool, United Kingdom
\item \Idef{org128}University of Science and Techonology of China, Hefei, China
\item \Idef{org129}University of South-Eastern Norway, Tonsberg, Norway
\item \Idef{org130}University of Tennessee, Knoxville, Tennessee, United States
\item \Idef{org131}University of the Witwatersrand, Johannesburg, South Africa
\item \Idef{org132}University of Tokyo, Tokyo, Japan
\item \Idef{org133}University of Tsukuba, Tsukuba, Japan
\item \Idef{org134}Universit\'{e} Clermont Auvergne, CNRS/IN2P3, LPC, Clermont-Ferrand, France
\item \Idef{org135}Universit\'{e} de Lyon, Universit\'{e} Lyon 1, CNRS/IN2P3, IPN-Lyon, Villeurbanne, Lyon, France
\item \Idef{org136}Universit\'{e} de Strasbourg, CNRS, IPHC UMR 7178, F-67000 Strasbourg, France, Strasbourg, France
\item \Idef{org137}Universit\'{e} Paris-Saclay Centre d'Etudes de Saclay (CEA), IRFU, D\'{e}partment de Physique Nucl\'{e}aire (DPhN), Saclay, France
\item \Idef{org138}Universit\`{a} degli Studi di Foggia, Foggia, Italy
\item \Idef{org139}Universit\`{a} degli Studi di Pavia, Pavia, Italy
\item \Idef{org140}Universit\`{a} di Brescia, Brescia, Italy
\item \Idef{org141}Variable Energy Cyclotron Centre, Homi Bhabha National Institute, Kolkata, India
\item \Idef{org142}Warsaw University of Technology, Warsaw, Poland
\item \Idef{org143}Wayne State University, Detroit, Michigan, United States
\item \Idef{org144}Westf\"{a}lische Wilhelms-Universit\"{a}t M\"{u}nster, Institut f\"{u}r Kernphysik, M\"{u}nster, Germany
\item \Idef{org145}Wigner Research Centre for Physics, Budapest, Hungary
\item \Idef{org146}Yale University, New Haven, Connecticut, United States
\item \Idef{org147}Yonsei University, Seoul, Republic of Korea
\end{Authlist}
\endgroup
\end{document}